\documentclass{emulateapj}
\usepackage{myaasmacros}

\usepackage{graphicx}
\usepackage{bm}
\usepackage{amsmath,amssymb}
\usepackage{amsfonts}
\usepackage{epsfig}

\def\ltap{\raisebox{-.55ex}{\rlap{$\sim$}} \raisebox{.4ex}{$<$}}
\def\gtap{\raisebox{-.55ex}{\rlap{$\sim$}} \raisebox{.4ex}{$>$}}
\def\gsim{\mathrel{\gtap}}
\def\lsim{\mathrel{\ltap}}

\shorttitle{Search for Anisotropy with TA}
\shortauthors{Telescope Array Collaboration}

\begin{document}

\title{Search for Anisotropy of Ultra-High Energy Cosmic Rays with
  the Telescope Array Experiment}

\iftrue 
\author{
T.~Abu-Zayyad$^{1}$, 
R.~Aida$^{2}$, 
M.~Allen$^{1}$, 
R.~Anderson$^{1}$, 
R.~Azuma$^{3}$, 
E.~Barcikowski$^{1}$, 
J.W.~Belz$^{1}$,
D.R.~Bergman$^{1}$, 
S.A.~Blake$^{1}$, 
R.~Cady$^{1}$
% ...
\\~
}
%
\affiliation{
$^1$University of Utah, High Energy Astrophysics Institute, Salt Lake City, Utah, USA\\
$^2$University of Yamanashi, Interdisciplinary Graduate School of Medicine and Engineering, Kofu, Yamanashi, Japan\\
$^3$Tokyo Institute of Technology, Meguro, Tokyo, Japan\\
% ...
}
\fi


\begin{abstract}
We study the anisotropy of Ultra-High Energy Cosmic Ray (UHECR) events
collected by the Telescope Array (TA) detector in the first 40 months 
of operation. Following earlier studies, we examine event sets
with energy thresholds of $10$~EeV, $40$~EeV, and $57$~EeV. We
find that the distributions of the events in right ascension and
declination are compatible with an isotropic distribution in all
three sets. We then compare with previously reported clustering of the
UHECR events at small angular scales. No significant clustering is
found in the TA data. We then check the events with $E>57$~EeV for
correlations with nearby active galactic nuclei. No significant
correlation is found. Finally, we examine all three sets for
correlations with the large-scale structure of the Universe. We find
that the two higher-energy sets are compatible with both an isotropic
distribution and the hypothesis that UHECR sources follow the matter
distribution of the Universe (the LSS hypothesis), while the event set
with $E>10$~EeV is compatible with isotropy and is not compatible with
the LSS hypothesis at 95\% CL unless large deflection angles are also 
assumed. We show that accounting for UHECR deflections in a realistic
model of the Galactic magnetic field can make this set compatible
with the LSS hypothesis. \\~~~~~~~~~~~~~~~~~~~~~~~~~~~~~~~~~~~~~~~
\end{abstract}

\maketitle

\section{Introduction}
\label{sec:introduction}

One of the keys to understanding the nature of the Ultra-High Energy
Cosmic Rays (UHECRs) is their distribution over the sky. This
distribution depends on the UHECR sources, as well on the UHECR
mass composition and large-scale magnetic fields, both Galactic
and extragalactic. Despite significant effort, none of these issues is
well understood at present.

Observation of the cutoff in the highest-energy part of the cosmic ray
spectrum \citep{Abbasi:2007sv,Abraham:2008ru} suggests that the UHECR
propagation at high energies is limited by the interaction with the cosmic
background radiation (the Greisen-Zatsepin-Kuzmin (GZK) effect
\citep{Greisen:1966jv,Zatsepin:1966jv}). One therefore expects that the
closest sources of UHECRs are situated within the GZK volume of the size
$\lsim 100$~Mpc.  At these scales the matter distribution in the Universe is
inhomogeneous, and so must be the distribution of the UHECR sources. If
propagation of UHECRs at these distances is quasi-rectilinear (whether or not
this is the case depends on both their composition and the magnetic
fields), one generally expects the UHECR flux to be anisotropic, showing
variations at large angular scales and possibly point sources.

If UHECR primary particles are protons, as suggested by the composition
measurements performed by the High Resolution Fly's Eye (HiRes) and 
the Telescope Array (TA) experiments
\citep{Abbasi:2009nf,Tameda:2010-uhecr2010}, the UHECR propagation is, in
fact, expected to be quasi-rectilinear. With the existing estimates of the
Galactic magnetic field \citep{Han2006,Sun:2007mx,Pshirkov:2011um} and bounds
on the extragalactic ones \citep{Kronberg:1993vk}, the deflections of protons
should be relatively small. For instance, a random extragalactic field of magnitude
1~nG and correlation length of $\sim 1$~Mpc would deflect a proton of energy
$10^{20}$~eV by about $2^\circ$ over a distance of 50~Mpc, while the Galactic
field would produce deflections of order $2-4^\circ$ depending on the
direction. In this case a sizeable anisotropy may be expected regardless of
the density of the UHECR sources down to energies as low as $10^{19}$~eV.

On the contrary, if the composition at highest energies is heavy or
predominantly heavy, as the results of the Pierre Auger Observatory
(PAO) \citep{Abraham:2010yv} seem to indicate, the quasi-rectilinear
propagation is not expected for the bulk of UHECRs. Some anisotropy at
large angles may still arise if the extragalactic fields are
sufficiently small and the density of sources is such that only a few
nearby ones contribute to the observed flux, but the small-scale
anisotropy would be suppressed (for recent analyses see, e.g.,
\citet{Giacinti:2010dk,Takami:2012uw}). Thus, the study of the
UHECR anisotropy may shed light on both the mass composition and the
density of the UHECR sources 
\citep{Dubovsky:2000gv,Yoshiguchi:2002rb,Yoshiguchi:2004np,Kachelriess:2004pc}.

Numerous attempts at detection of the UHECR anisotropy have been made
previously. Early studies indicated clustering of the UHECR events at
small angular scales \citep{Hayashida:1996bc,Tinyakov:2001ic}. On the
basis of small-scale correlations, different classes of putative
sources of UHECR were suggested (see, e.g.,
\citet{Gorbunov:2004bs,Abbasi:2005qy,Cronin:2007zz,Abraham:2007si}). More
recently, the Pierre Auger Observatory  has claimed correlations
of UHECRs with the nearby Active Galactic Nuclei (AGN)
\citep{Cronin:2007zz,Abraham:2007si} which were not confirmed by
observations in the Northern hemisphere \citep{Abbasi:2008md}.

At larger angular scales, evidence for correlations with the
supergalactic plane was claimed by \citet{Stanev:1995my},
\citet{Glushkov:2001jm}, and \citet{Glushkov:2001kb} but not confirmed
by other authors
\citep{Hayashida:1996bc,Kewley:1996zt,Bird:1998nu}. Also,
\citet{Kashti:2008bw} found the anisotropy in the PAO data 
which was not confirmed by the HiRes
data in the Northern hemisphere \citep{Abbasi:2010xt}.

In this paper we present the anisotropy analysis of UHECR observed by
the Surface Detector (SD) of the Telescope Array in the first $40$ months 
of its operation. TA is a hybrid UHECR detector located in the
Northern hemisphere in Utah, USA (39$^\circ$17$'$48$''$ N,
112$^\circ$54$'$31$''$ W) which has been fully operational since
March 2008. It consists of 507 scintillator detectors covering the
area of approximately 700~km$^2$ (for details see
\citet{AbuZayyad:2012kk}). The atmosphere over the surface array is
viewed by 38 fluorescence telescopes arranged in 3 stations (see
\citet{Tokuno:2012mi}). The surface detector of TA is the largest in
the Northern hemisphere.

In this paper we focus on testing previous
observations.  Namely, we consider the clustering of the UHECR events at small
angular scales (as would be produced by bright point sources), possible
correlation with nearby AGN and correlation of the TA events with the
large-scale structure (LSS) of the Universe. Following previous analyses, we
consider three {\em a priori} chosen energy thresholds: $10$~EeV, $40$~EeV, and
$57$~EeV. It should be noted that different experiments may have different
energy scales due to different systematic errors in the energy determination,
which may affect the selection of the events. When referring to the results of
other experiments, we assume the energy scales as reported by these
experiments. In statistical tests which require a pre-defined confidence level
we set the latter to 95\%.

The paper is organized as follows. In Section~\ref{sec:data} we describe the
data sets used. In Section~\ref{sec:autoc-funct} we examine the event sets for a
presence of small-scale clustering by studying the UHECR auto-correlation
function. In Section~\ref{sec:corr-with-point} we consider correlations of UHECR
events with nearby AGNs. Section~\ref{sec:correlation-with-lss} describes our
search for correlations of the UHECR events with the large-scale structure of
the Universe. In Section~\ref{sec:conclusions} we summarize the results and
present conclusions.
\vskip 30pt 

\section{Data}
\label{sec:data}

Among the existing TA data sets (SD data, Fluorescence Detector (FD)
data in mono and stereo modes, and hybrid detector data) the SD data set has by
far the largest number of events. 
The present analysis is based on the data collected in the period
2008.05.11--2011.09.15 (40 months) of operation by the TA 
surface detector array.  
Cutting events with zenith angle $> 45^\circ$, the
SD data set contains 988 events with energies $> 10$~EeV, 57 events
with $E>40$~EeV, and 25 events with $E>57$~EeV. 
This is the largest UHECR set to date in the Northern hemisphere.

The angular resolution of TA events with $E>10$~EeV, is approximately 
$1.5^\circ$.
This follows from the comparison of the thrown and reconstructed arrival
directions of simulated data sets, and is supported by the direct comparison
between the SD and FD arrival directions of hybrid events.  The energy
resolution of the TA surface detector at $E>10$~EeV is better than 20\%
\citep{AbuZayyad:2012ru}.

The exposure of the TA surface array is calculated by the Monte-Carlo
technique with full simulation of the detector, which will be described
elsewhere. As follows from the Monte-Carlo simulations, the acceptance of the
TA surface detector for $E>10$~EeV and zenith angle cut of $45^\circ$
is close to the geometrical one. For reasons of computational efficiency, in
the present analysis aimed at anisotropy at relatively small angular scales we
use the geometrical acceptance to generate random event sets.

\begin{figure}
\begin{picture}(220,250)(0,0)
\put(0,170){\includegraphics[height=80pt]{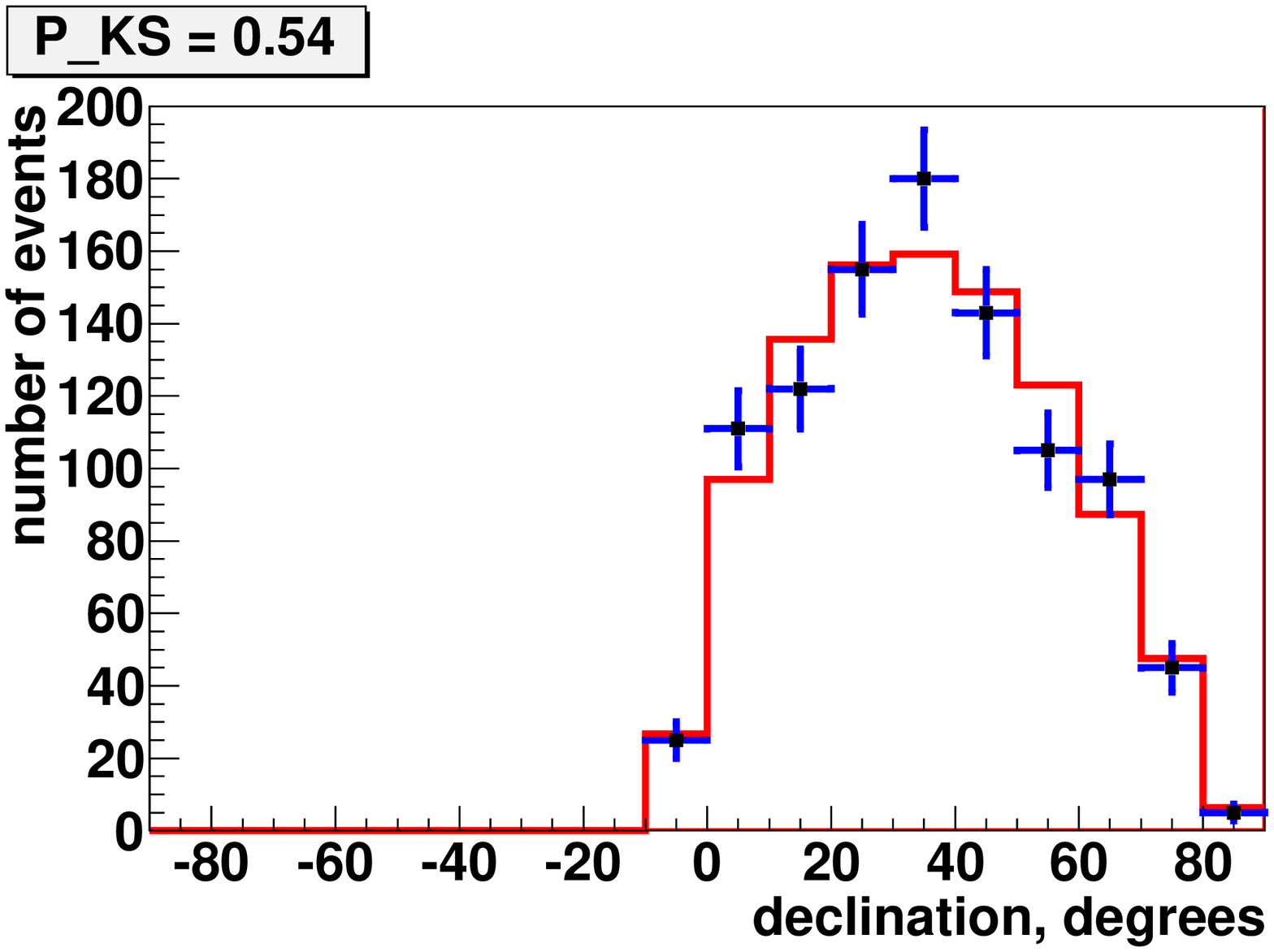}}
\put(125,170){\includegraphics[height=80pt]{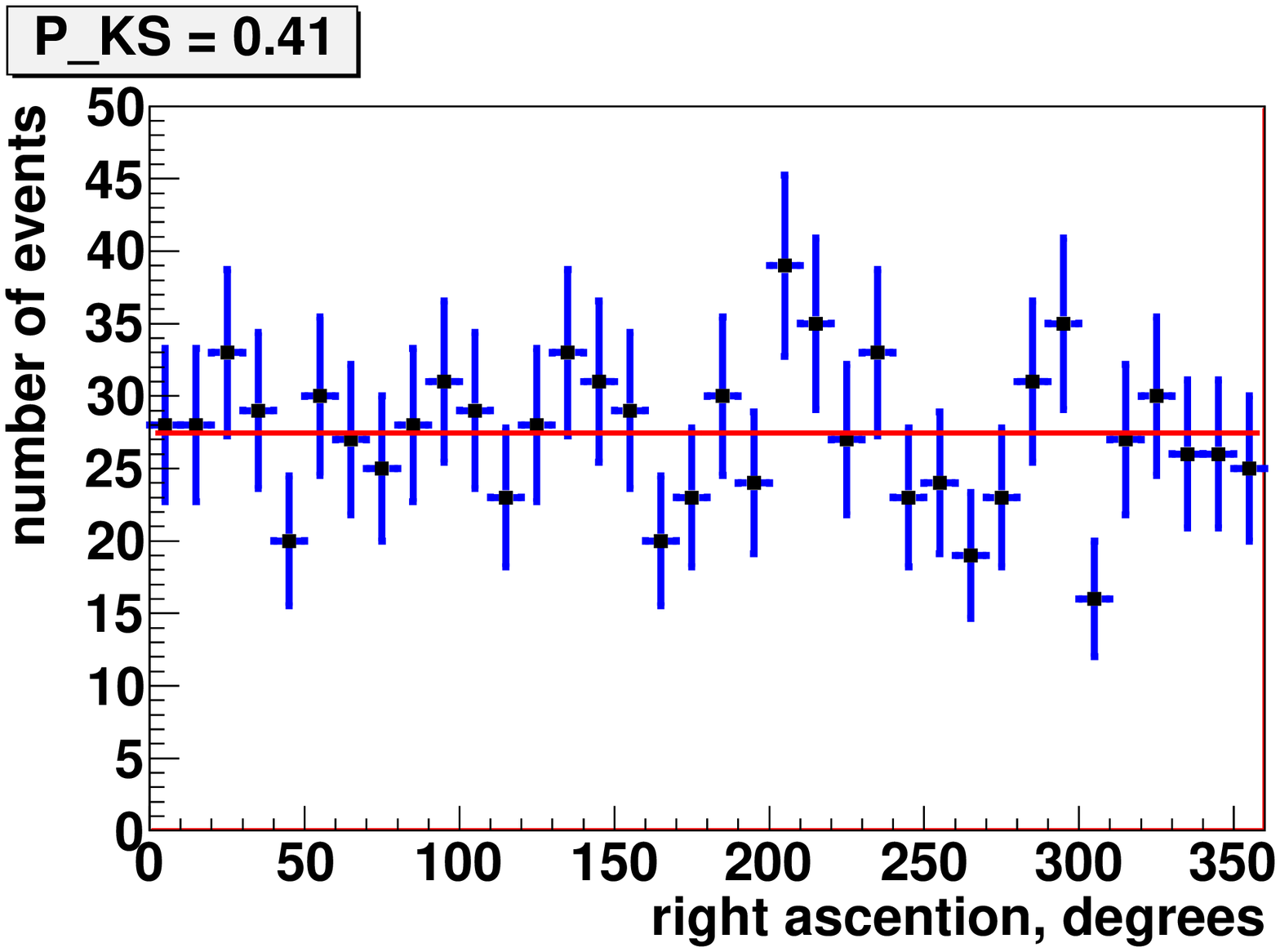}}
\put(0,85){\includegraphics[height=80pt]{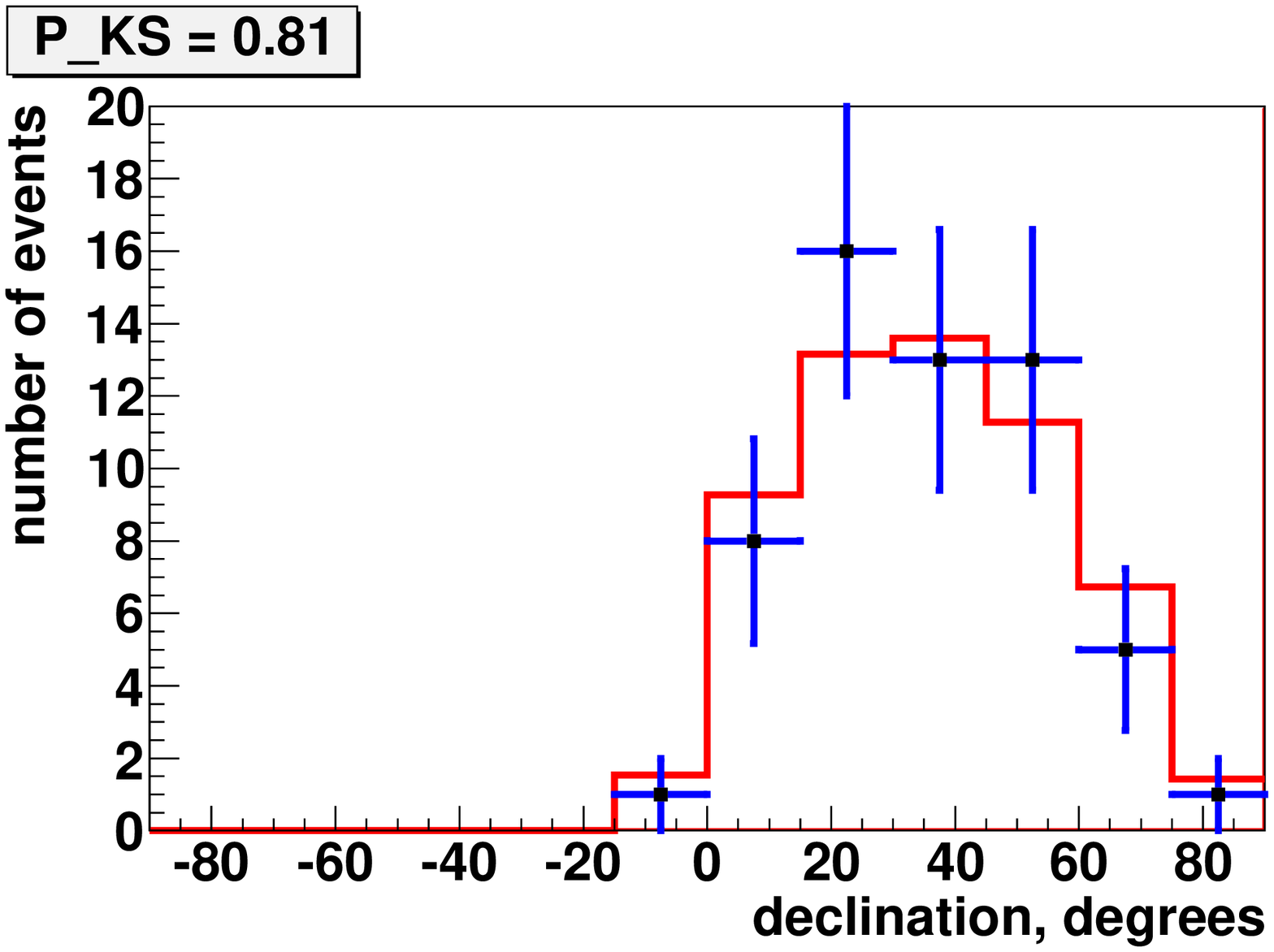}}
\put(125,85){\includegraphics[height=80pt]{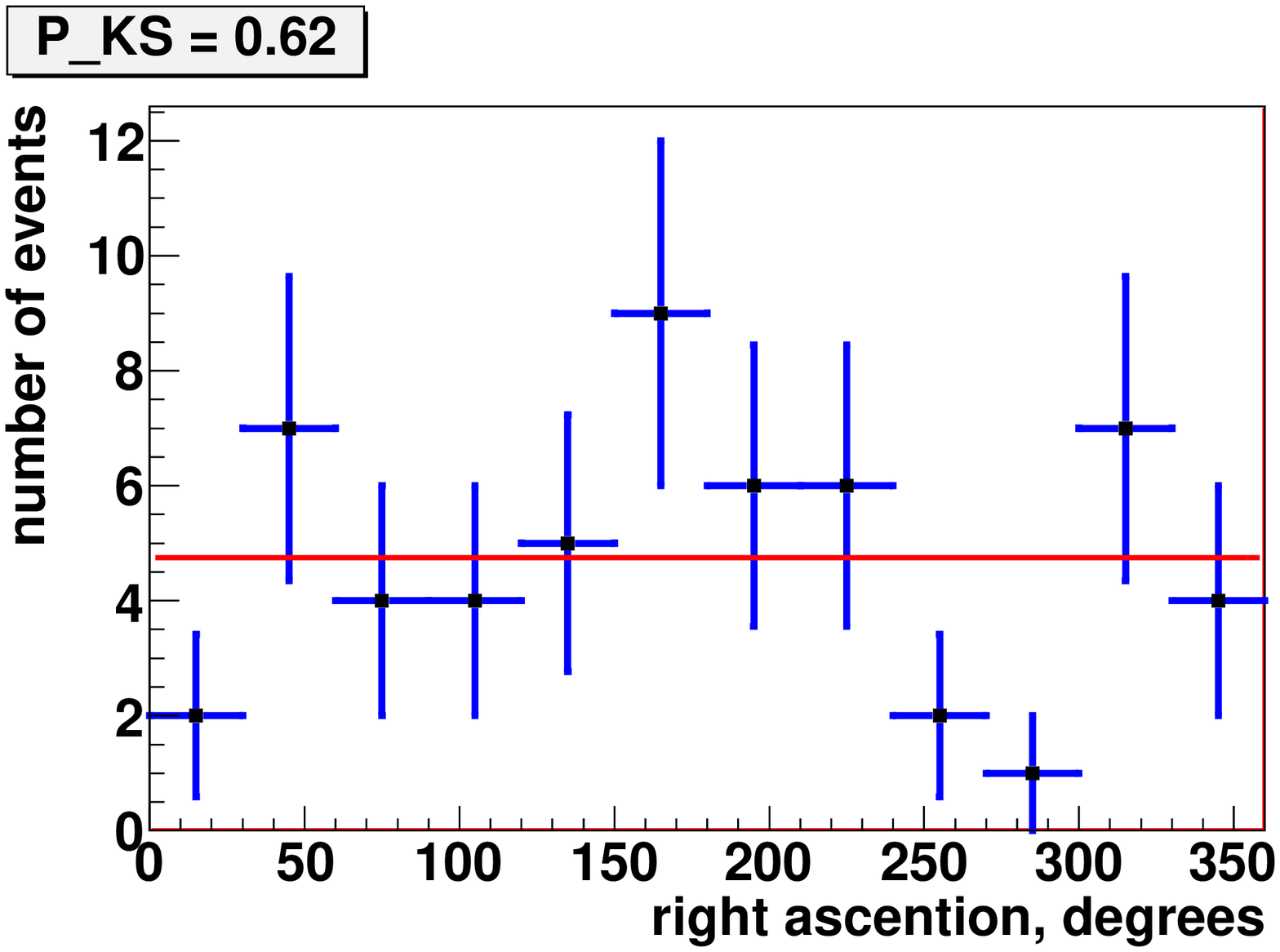}}
\put(0,0){\includegraphics[height=80pt]{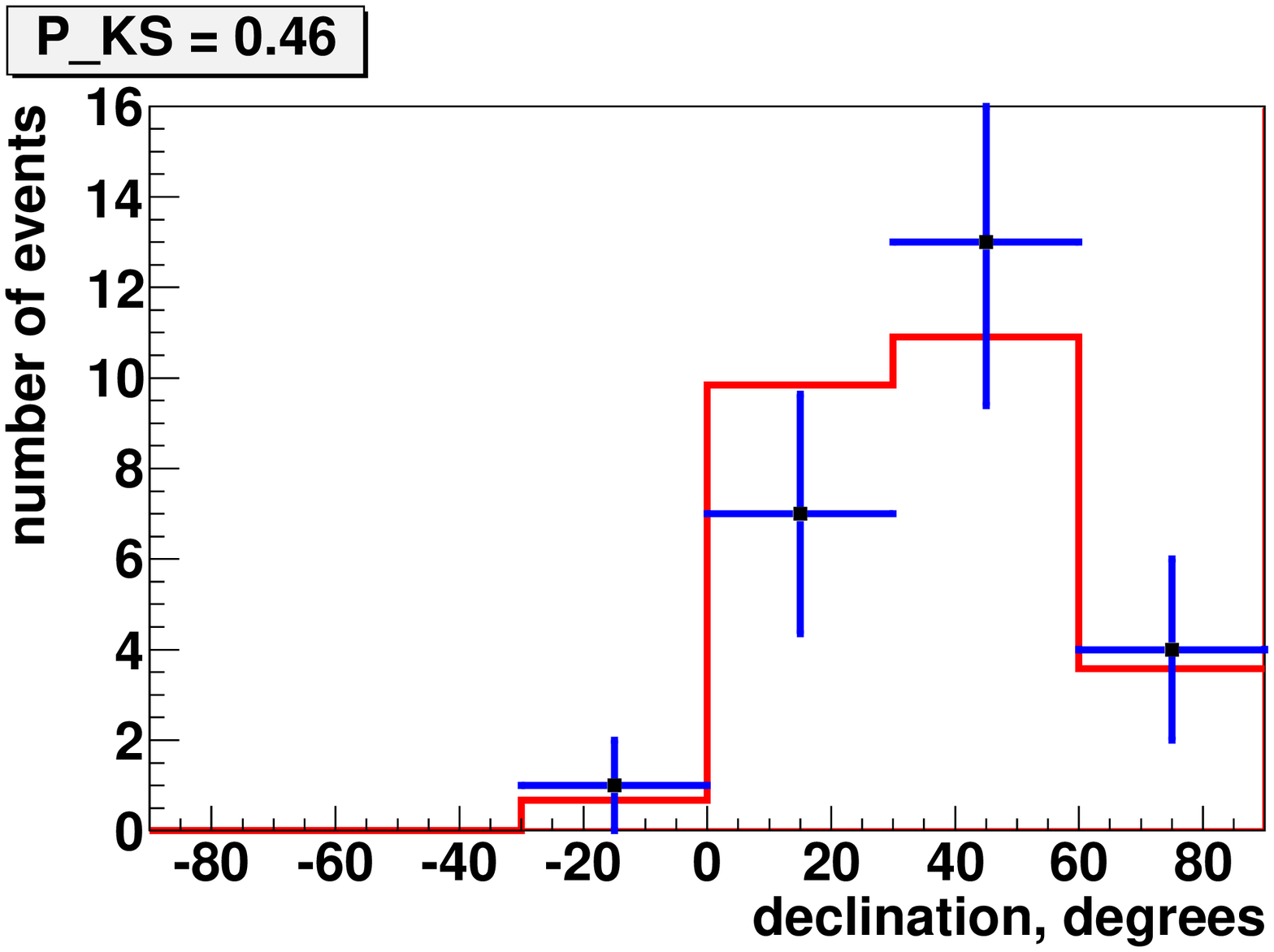}}
\put(125,0){\includegraphics[height=80pt]{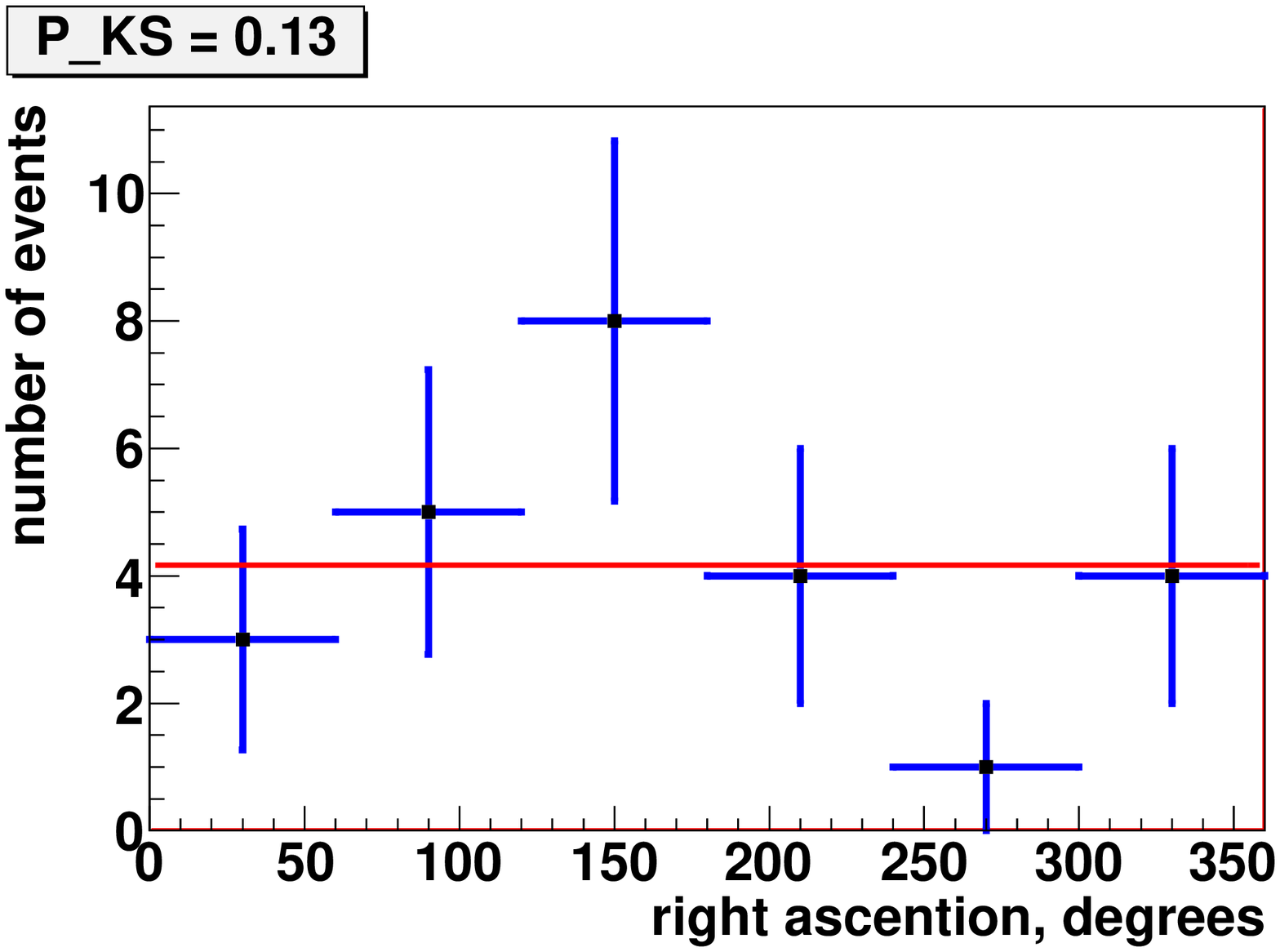}}
\end{picture}
  \caption{\label{fig:geom-exposure} Comparison between the data (blue
    points) and the sets of $10^4$ events simulated with the geometrical
    exposure (red histogram) at energies $10$~EeV, $40$~EeV, and $57$~EeV
    (top, middle, and bottom rows, respectively). Plots show the
    distribution of events in declination (left column) and right
    ascension (right column). The compatibility of the two
    distributions by the Kolmogorov-Smirnov test is given as $P_{\rm
      KS}$ in the upper left corner of each plot. }
\end{figure}

Figure~\ref{fig:geom-exposure} shows the comparison between the distributions in
declination (left column) and right ascension (right column) of the events
simulated with the geometrical exposure (red line) and the data (blue data
points) at the energy thresholds of $10$~EeV, $40$~EeV, and $57$~EeV (top,
middle, and bottom rows, respectively). The compatibility of expected and
observed distributions in all 6 cases was checked by the Kolmogorov-Smirnov
(KS) test. The lowest KS probability was $p=0.13$ for the distribution in the
right ascension at $E>57$~EeV. Thus, all three sets are compatible with a
uniform distribution.

\section{Autocorrelation function}
\label{sec:autoc-funct}

The AGASA experiment reported clustering of UHECR events with
$E>40$~EeV at the angular scale of $2.5^\circ$ \citep{Hayashida:1996bc}. 
Here we repeat this analysis using the TA data set. 

The procedure is as follows: for a given angular separation, $\delta$, we count
the number of pairs of observed events that are separated by an angular
distance less than $\delta$, thus obtaining the data count.  We then generate
a large number (typically, $10^5$) of Monte-Carlo (MC) event sets each having
the same number of events as the real data set. The simulated sets are generated
with a uniform distribution according to the TA exposure. In each MC set we
count pairs of events in the same way as in the data, which gives the MC count
for that set. We then calculate the average MC count for all of the MC sets. 
This represents the expected number of pairs for the angular scale $\delta$, 
assuming a uniform cosmic ray distribution. 
For each value of $\delta$ we then determine the
fraction of simulated sets where the number of pairs is greater than or equal
to the number of pairs in the data. This gives the $p$-value, $P(\delta)$, that
describes how likely the excess of pairs, if found in the data, is to occur as
a result of a fluctuation in a random set. 
Small values of $P(\delta)$, thus indicate a departure from 
uniformity at the corresponding angular scale.

We first perform a blind test of the AGASA claim. Fixing the energy
threshold to $40$~EeV and the separation angle to $\delta = 2.5^\circ$ we find
0 pairs while 1.5 pairs are expected in the case of a uniform
distribution. Therefore, there is no excess of small-scale clusters in the TA
data.

\begin{figure}
\begin{picture}(220,250)(0,0)
\put(0,170){\includegraphics[height=80pt]{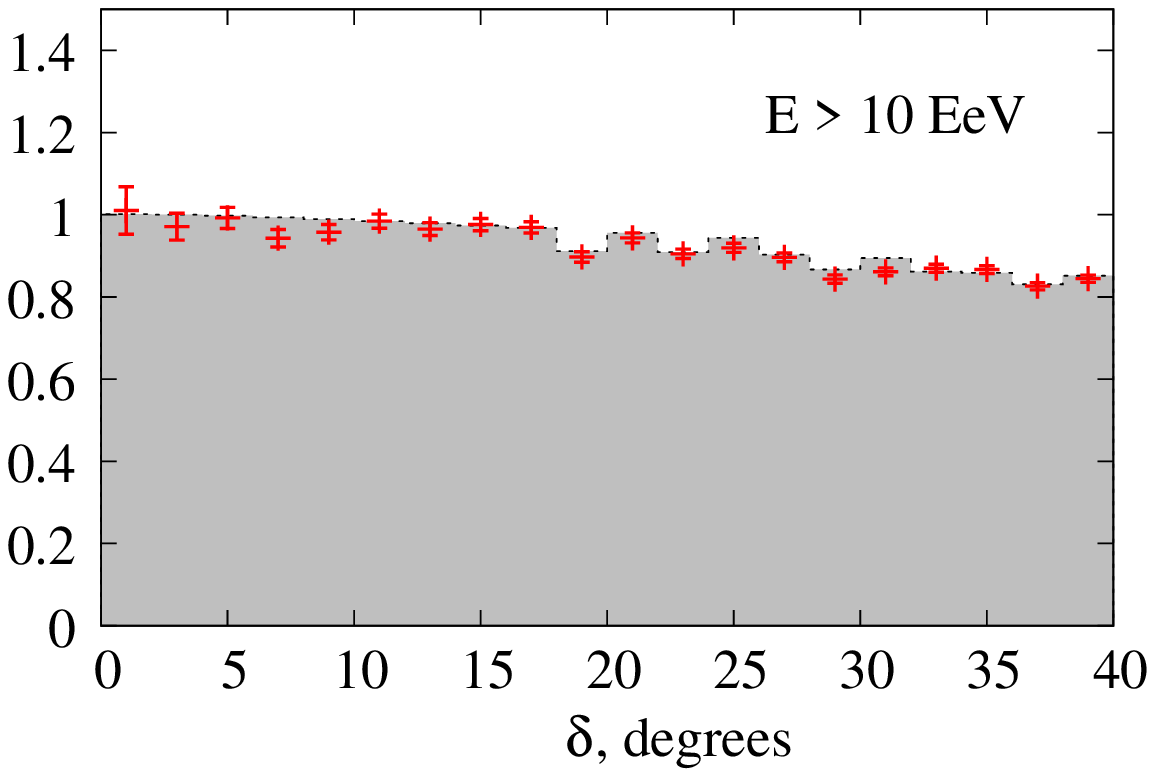}}
\put(125,170){\includegraphics[height=80pt]{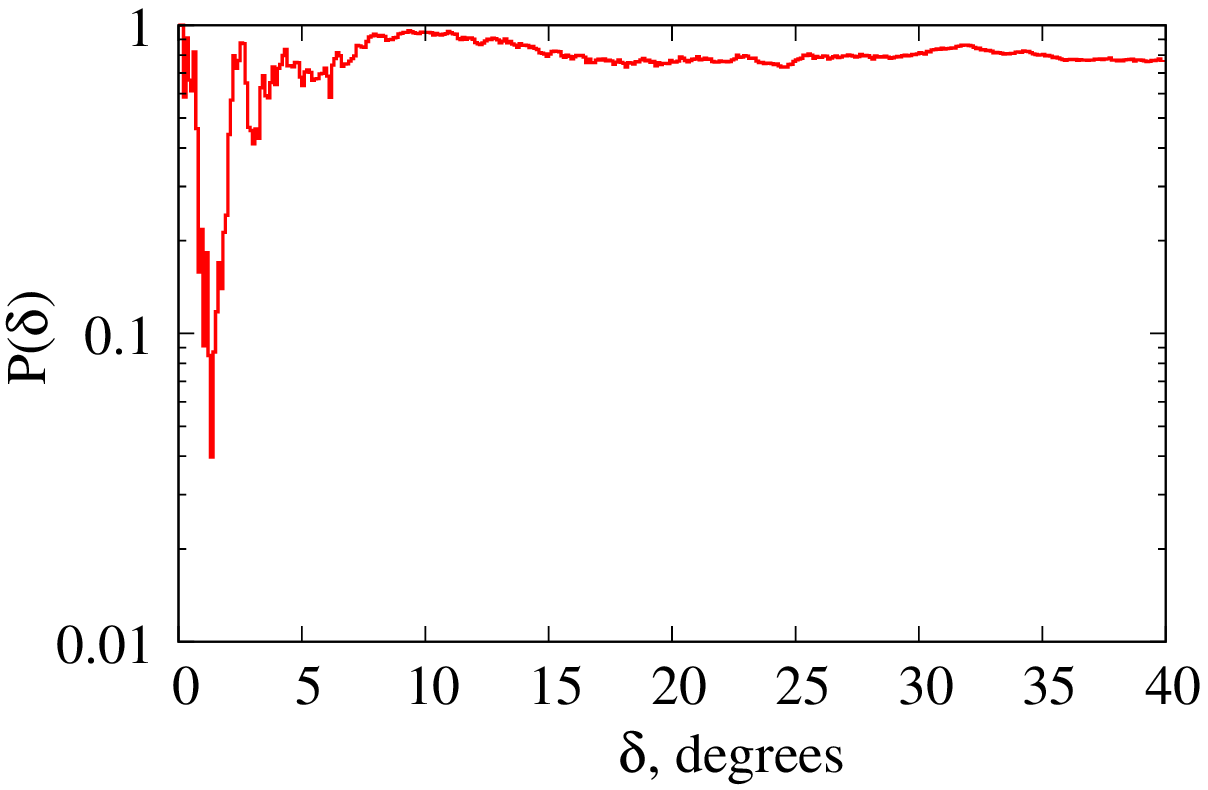}}
\put(0,85){\includegraphics[height=80pt]{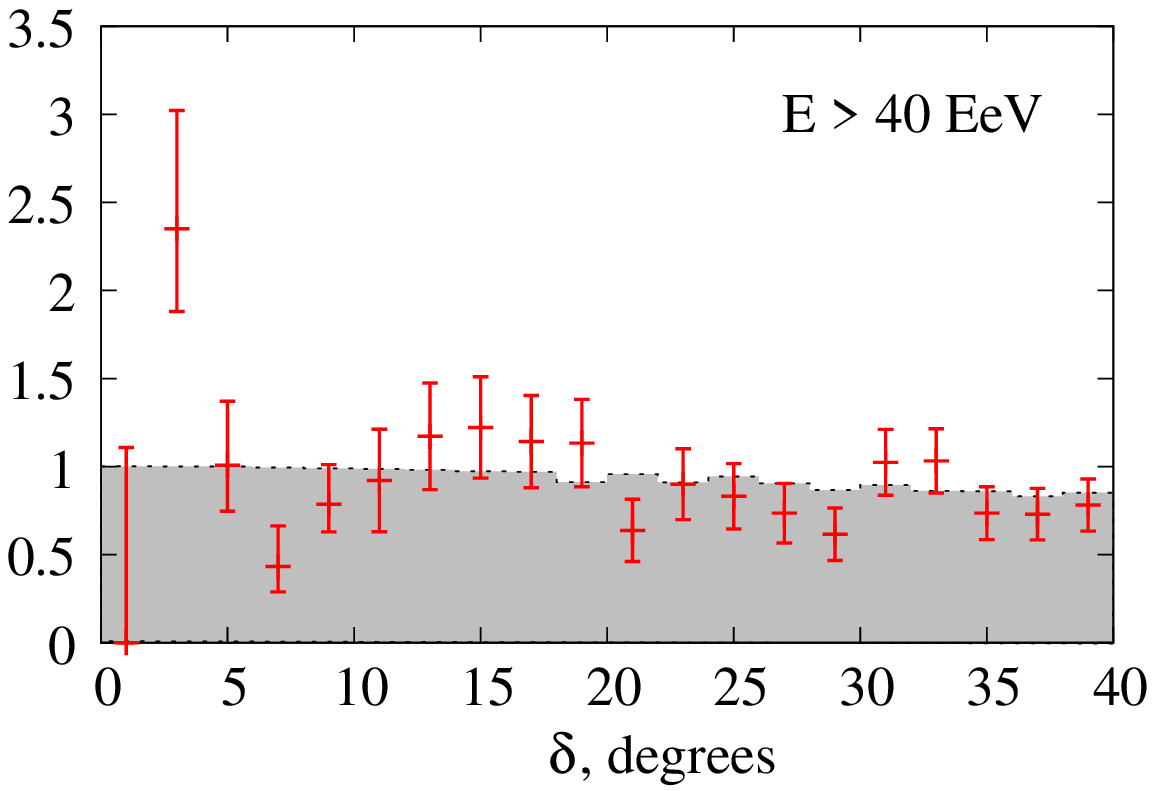}}
\put(125,85){\includegraphics[height=80pt]{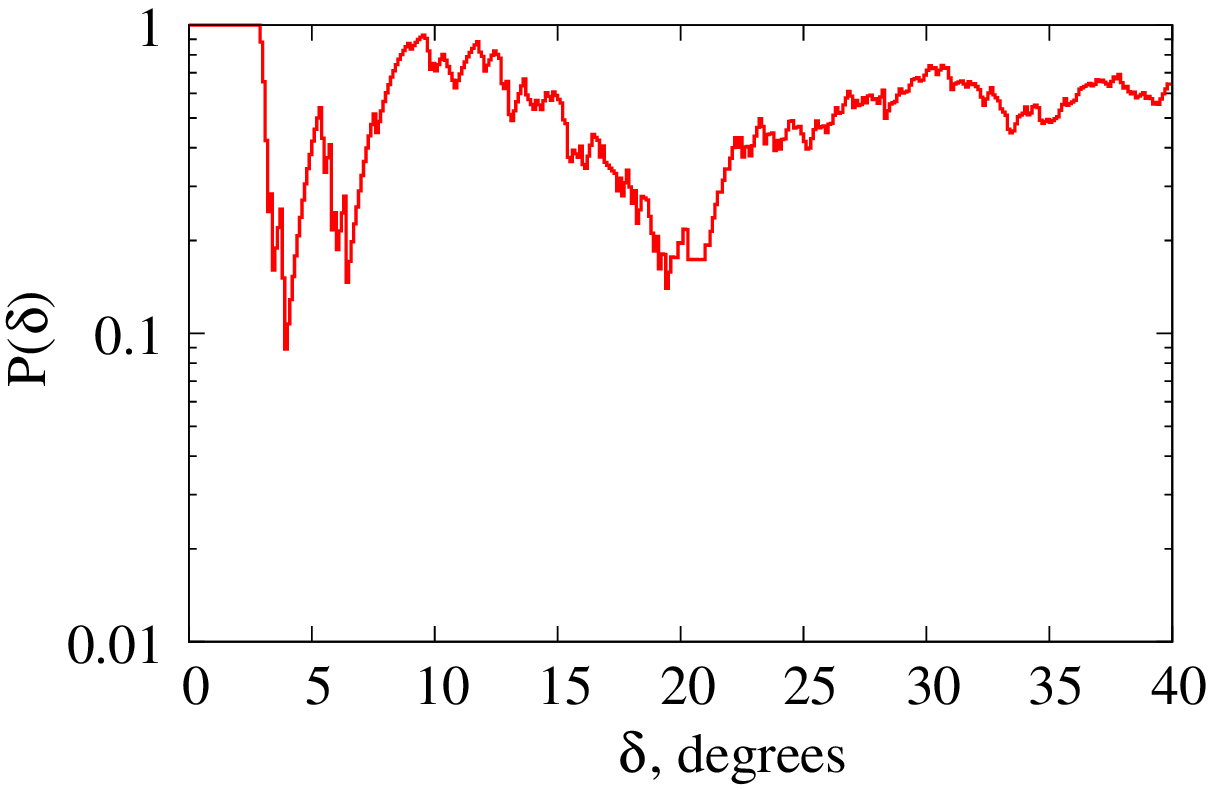}}
\put(0,0){\includegraphics[height=80pt]{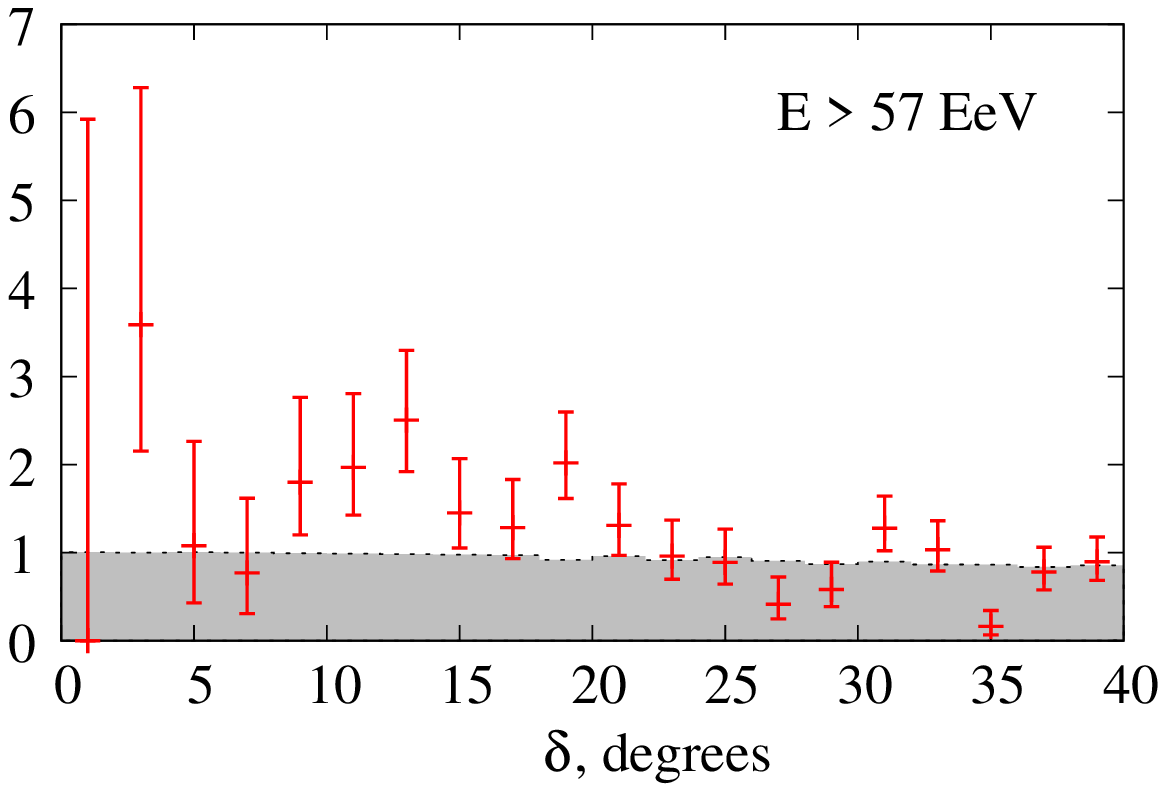}}
\put(125,0){\includegraphics[height=80pt]{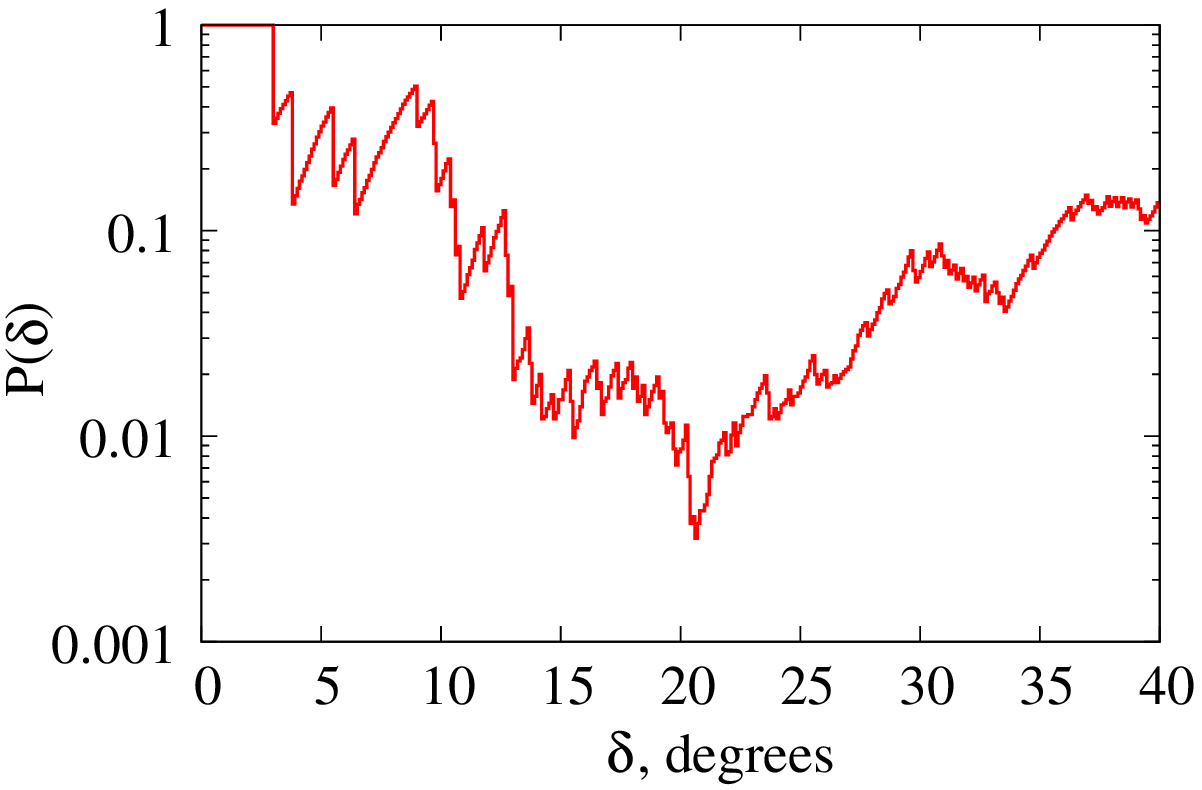}}
\end{picture}
  \caption{\label{fig:auto} Autocorrelations in the TA data sets at
    $E>10$~EeV, $E>40$~EeV, and $E>57$~EeV (top, middle, and bottom rows,
    respectively).  {\em Left panels:} the number of pairs with angular
    separations $\delta$ normalized to the area of the angular bin (data
    points), compared to the expectation for the uniform distribution (shaded
    histogram). The errors are 1-sigma Poisson errors. {\em Right panels:}
    probability, $P(\delta)$, that the excess of pairs with the angular
    separation less than $\delta$ occurs as a fluctuation in a uniform
    distribution. Small $P(\delta)$ indicates a departure from isotropy. }
\end{figure}

We next extend the analysis to all angular scales. No significant
excess is found.  The results are illustrated in Figure~\ref{fig:auto}
for angles from $0$ to $40^\circ$ and three energy thresholds of
$10$~EeV, $40$~EeV, and $57$~EeV as specified on the plots. For each
energy threshold, the left panel shows the number of pairs with the
angular separations $\delta$ binned in $2^\circ$ bins (data
points). The shaded region represents the average number of pairs
expected in the case of the uniform distribution. Both the data and
the uniform expectation are normalized bin-by-bin to the area of the
bin, so that in the case of a uniform full-sky exposure the
expectation would be flat. The overall normalization is set in such a
way that the expectation in the first bin equals one.

The right panels of Figure~\ref{fig:auto} show the dependence of the
p-value, $P(\delta)$, on the separation angle, $\delta$, for the
corresponding energy.  Note that $P(\delta)$ is a cumulative quantity
since it takes into account all the pairs separated by angles from 0
to $\delta$.  For this reason a small, but coherent over several bins,
excess at angles from 10 to 20 degrees on the lower left panel of
Figure~\ref{fig:auto} produces a more significant feature in the
corresponding $P(\delta)$, lower right panel of
Figure~\ref{fig:auto}. This feature corresponds to the group of events
visible on the sky map (see the lower panel of
Figure~\ref{fig:p-skymaps}).

When accessing the significance of departures from isotropy on the
basis of $P(\delta)$ represented in Figure~\ref{fig:auto}, one should
take into account the fact that the angular scale of the excess is not
known in advance. Thus, there is a statistical penalty for choosing
this scale {\em a posteriori} (see \citet{Tinyakov:2004bb} for a
detailed discussion). Taking this penalty into account, none of the
three examined data sets shows a significant deviation from an
isotropic distribution.

Interestingly, although close clusters in the high-energy TA event set are
absent, one of the TA events falls within $1.7^\circ$ of 
a high energy event observed by the Auger Observatory \citep{Abreu:2010zzj}.
Both events have $E>10^{20}$~EeV. 
The center of the doublet has the Galactic coordinates $l=36^\circ$, $b=-4.3^\circ$. 

\section{Correlation with Active Galactic Nuclei}
\label{sec:corr-with-point}

The Auger collaboration has reported a correlation
\citep{Cronin:2007zz,Abraham:2007si} between UHECRs with $E>57$~EeV
and the nearby (redshift $z\leq 0.018$ or, equivalently, distance
$d<75$~Mpc) Active Galactic Nuclei (AGNs) from the Veron-Cetty \&
Veron (VCV) catalog \citep{VeronCetty:2006zz}. The greatest
correlation was observed at the angle of $3.1^\circ$. In the control
data set, the number of correlating events was 9 out of 13, which
corresponds to about 69\% of events.  The Auger collaboration has
recently updated the analysis and found that a smaller fraction of the
UHECR events correlates with the same set of AGNs in the latest UHECR
data set \citep{Abreu:2010zzj} than in the original one. Out of 55 
events with $E>55$~EeV, 21 were found to correlate with AGNs, which
corresponds to a fraction of correlating events equal to 38\%.  In
this section we test the TA data for correlations with AGN.

The set of 472 nearby AGNs used by \citet{Cronin:2007zz} contains 7
objects listed at zero redshift, all in the field of view of TA. Of
these 7 objects, two are stars, one is a quasar with unknown redshift,
one is a Seyfert 2 galaxy, two are spiral galaxies (including the
Andromeda galaxy) and one is a dwarf spheroidal galaxy. We exclude
these objects from the analysis, which leaves 465 objects in the AGN
catalog.

The TA exposure is peaked in the Northern hemisphere, so that the AGNs
visible to TA are largely different from those visible to Auger,
though there is some overlap.  The distribution of nearby AGNs over
the sky is not uniform because of the large scale structure (see
Section~\ref{sec:correlation-with-lss} for more detail) and because
the VCV catalog is not complete: due to observational bias it tends to
contain more objects in the Northern hemisphere. For this reason, a
larger fraction of events is expected to correlate with AGNs in the TA
data under the assumption that AGNs are sources of the observed
UHECRs.  Taking into account the distribution of nearby AGNs over the
sky and assuming equal AGN luminosities in UHECR, we estimated the
correlating fraction will be $\sim 73$\% for TA on the basis of the
original PAO claim, and $\sim 43$\% on the basis of the updated
analysis by PAO.

\begin{figure}
 \begin{picture}(220,150)(0,0)
\put(0,20){ \includegraphics[height=.16\textheight]{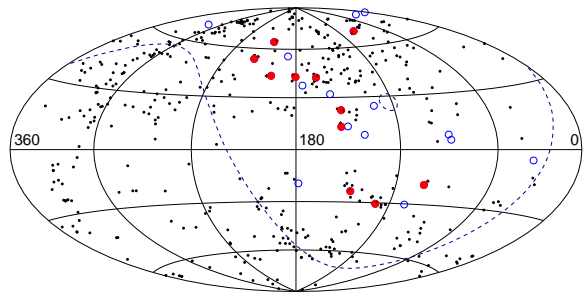} }
 \end{picture}
  \caption{\label{fig:E57vsAGN} Hammer projection of the TA cosmic ray events
with $E>57$~EeV and nearby AGNs in the Galactic coordinates. Correlating and
non-correlating events are shown by filled red and empty blue circles,
respectively. AGNs are represented by black dots. The dashed line
shows the boundary of the TA exposure. }
\end{figure}

The sky map of TA events with $E>57$~EeV and nearby AGNs from the VCV
catalog is represented in Figure~\ref{fig:E57vsAGN} in Galactic coordinates. The
cosmic rays are shown by filled red (correlating events) and empty blue circles
(non-correlating events). AGNs are shown by black dots.

\begin{figure}
 \begin{picture}(220,150)(0,0)
\put(0,10){ 
  \includegraphics[height=.2\textheight]{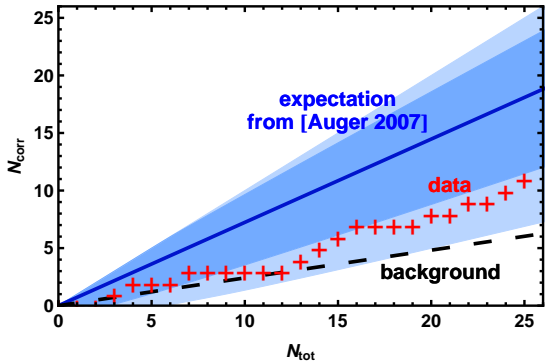} }
 \end{picture}
  \caption{\label{fig:AGNsignificance} The number of TA events with
    $E>57$~EeV correlating with VCV AGNs as a function of the total
    number of events. The expectation according to the original PAO
    claim is represented by the blue line together with the 1- and 2-sigma
    significance bands. The black dashed line shows the expected number of
    random coincidences. }
\end{figure}

Figure~\ref{fig:AGNsignificance} shows the number of TA events correlating with
AGNs as a function of the total number of events with $E>57$~EeV ordered
according to arrival time. The black dashed line represents the expected number
of random coincidences in case of a uniform distribution calculated via
Monte-Carlo simulation. The blue line shows the expected number of
correlating events as derived from the original PAO claim.  Shaded regions
represent 68\% and 95\% CL deviations from this expectation calculated by the
maximum likelihood method of Ref.~\citep{Gorbunov:2005fi}.  As is seen from
Figure~\ref{fig:AGNsignificance}, present TA data are compatible with both
isotropic distribution and the AGN hypothesis.

In the full TA SD data set, there are 11 correlating events out of 25
total, while the expected number of random coincidences for this total
number of events is 5.9. Making use of the binomial distribution with
the probability of a single event to correlate $p_{\rm
  iso}=0.24$, one finds that such an excess has
probability of $\sim 2\%$ to occur by chance with isotropic
distribution of arrival directions.

\section{Correlation with LSS}
\label{sec:correlation-with-lss}

Even though the sources of UHECRs are not known, their distribution in
space at large scales must follow that of the ordinary matter. The
latter is anisotropic at scales below $\sim 100$~Mpc forming the
Large-Scale Structure (LSS) of the Universe that consists of galaxy
clusters, filaments and voids. If UHECRs are not strongly deflected on
their way to Earth, their distribution over the sky should correlate
with the nearby structures, with over-densities corresponding to close
clusters and under-densities corresponding to voids.

The amplitude of anisotropy depends on the UHECR propagation length (the
larger is the propagation length, the smaller contributions of the local
structures and, therefore, the anisotropy) and on the UHECR deflections.  
In this section the propagation of UHECR is calculated assuming they are
protons. However, it should be noted that regardless of whether the UHECR
composition is heavy or light, their propagation length changes with energy
roughly in the same way and becomes of order several tens of megaparsecs as
the energy approaches $10^{20}$~eV. Thus, the most important parameter that
determines the amplitude of the anisotropy at a given energy is the typical
deflection angle which we denote as $\theta$ (which is, of course, very
different for heavy and light composition).

The goal of this analysis is to determine which values of $\theta$ are
compatible with the space distribution of the TA events. In principle, this
can be done at all energies. To minimize statistical penalties, we limit our
analysis to the energy thresholds of $10$~EeV, $40$~EeV, and $57$~EeV.

\subsection{Statistical Method}
\label{sec:statistical-method}

To test the compatibility between the observed UHECR distribution over the sky
and that expected under the LSS hypothesis (that is, the hypothesis that UHECR
sources trace matter distribution in the Universe), we employ the method
developed by \citet{Koers:2008ba} and used previously in the analysis of the
HiRes data \citep{Abbasi:2010xt}. In this method, one first computes the UHECR
flux distribution expected under the LSS hypothesis and then compares it to 
the observed one by the flux sampling test.

The matter distribution in the nearby Universe may be inferred from the
complete galaxy catalogs containing the redshift information. In this work we
use the 2MASS Galaxy Redshift Catalog (XSCz)\footnote{We are grateful to
  T.~Jarrett for providing us with the preliminary version of this catalog.}
that is derived from the 2MASS Extended Source Catalog (XSC), with redshifts
that have either been spectroscopically measured (for most of the objects) or
derived from the 2MASS photometric measurements. This catalog provides the
most accurate information about 3D galaxy distribution to date.

For the flux calculations, we use the flux-limited subsample of galaxies with
apparent magnitude $m\leq 12.5$. For fainter objects, the completeness of
the catalog degrades progressively, while their inclusion does not change the
results considerably. We exclude objects closer than $5$~Mpc in order to
avoid breaking of the statistical description (if such objects are assumed
to be sources of UHECR, they have to be treated individually). We also cut out
galaxies at distances further than $250$~Mpc replacing their combined
contribution by a uniform flux normalized in such a way that it provides the
correct fraction of events as calculated in the approximation of a uniform
source distribution. The quantitative justification of these procedures can be
found in \citep{Koers:2009pd}. The resulting catalog contains 106\,218
galaxies, which is sufficient to accurately describe the flux distribution at
angular scales down to $\sim 2^\circ$. The UHECR flux distribution 
is reconstructed from this flux-limited catalog by the weighting 
method proposed by \citet{Lynden-Bell:1971} and adapted to flux calculations
by \cite{Koers:2009pd}. 

The XSCz catalog loses completeness in the band of roughly $\pm 10^\circ$
around the Galactic plane and especially around the Galactic center. The size
of this region is not much larger than a typical deflection of a proton even
at $57$~EeV, so this gap may be bridged without loss of accuracy. Away from
the Galactic center at $|l|>60^\circ$ where only a fraction of the galaxies
(the dimmer part) is missing in the catalog, we apply a $l-$ and $b-$dependent
weight correction to the remaining galaxies so as to compensate for the
missing ones. In the region close to the Galactic center, $|l|<60^\circ$, we
extrapolate the flux density from the adjacent regions in a straightforward
manner. The latter is not an accurate procedure; however, the Galactic center
region overlaps with the TA exposure only slightly, and this inaccuracy is not
important for our results as can be checked by excluding this region from the
analysis.

\begin{figure}
\includegraphics[width=.95\columnwidth]{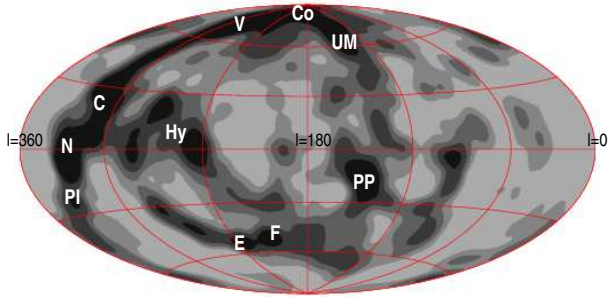}
\caption{Sky map of expected flux at $E>57$~EeV (Galactic coordinates). The
  smearing angle is $6^\circ$. Letters indicate the nearby structures as
  follows: {\bf C}: Centaurus supercluster (60 Mpc); {\bf Co}: Coma cluster
  (90 Mpc); {\bf E}: Eridanus cluster (30 Mpc); {\bf F}: Fornax cluster (20
  Mpc); {\bf Hy}: Hydra supercluster (50 Mpc); {\bf N}: Norma supercluster (65
  Mpc); {\bf PI}: Pavo-Indus supercluster (70 Mpc); {\bf PP}: Perseus-Pisces
  supercluster (70 Mpc); {\bf UM}: Ursa Major (20 Mpc); {\bf V}: Virgo cluster
  (20 Mpc).}
\label{fig:E57skymap}
\end{figure}

When propagating the UHECR primary particles from a source to the
Earth, we assume them to be protons and take full account of the
attenuation processes.  The injection index at the source is taken to
be 2.4, which is compatible with the UHECR spectrum observed by HiRes
and TA \citep{AbuZayyad:2012ru} assuming proton composition and the source evolution parameter
$m=4$ \citep{Gelmini:2007jy}. 
We also assume that the effects of both the Galactic
and extragalactic magnetic fields can be approximated by a single parameter,
the Gaussian smearing angle $\theta$. We consider $\theta$ a free parameter
and vary it in the range $2 - 20^\circ$. In general, the deflections of UHECR
in magnetic fields contain both random and regular parts, the latter being due
to the regular component of the Galactic magnetic field. The regular
deflections are not Gaussian. However, the statistical test we use here is not
sensitive to the coherent character of deflections provided they do not exceed
$10-20^\circ$ as set by the typical size of the flux variations due to local
structures (cf. Figure~\ref{fig:E57skymap}). So, for the most part of the
analysis we will use the Gaussian smearing to represent all the deflections
without making the distinction between the regular and random ones. Later, in
Section~\ref{sec:acco-galact-magn} we will discuss, in the case of the lowest
energy set and the largest deflections, the effect of explicitly accounting
for the regular component of the Galactic magnetic field.

To calculate the expected flux, we assume that UHECR sources follow the space
distribution of galaxies. The simplest way to realize this assumption in
practice is to assign each galaxy an equal luminosity in UHECR's.  This is a
good approximation if the density of the UHECR sources is sufficiently high
(so that many sources are present in local structures contributing to the
anisotropy).  The contribution of each galaxy to the total flux is then
calculated taking into account the distance of the source and the
corresponding flux attenuation.  Individual contributions are smeared with the
Gaussian width $\theta$, so that the flux at a given point of the sky is a sum
of contributions of all the galaxies within the angular distance of order
$\theta$. Further details on the flux calculation can be found in
References~\citep{Koers:2008ba,Koers:2009pd,Abbasi:2010xt}.

Figure~\ref{fig:E57skymap} shows the flux map calculated by the above
procedure for an energy threshold of $57$~EeV and smearing angle
$\theta = 6^\circ$, not yet modulated with the TA exposure. Darker regions
correspond to higher flux. A band of each color integrates to $1/5$ of
the total flux.  One can identify the nearby structures which are
marked by letters on the plot as explained in the caption.

The next step is to compare the calculated flux distribution to the actual
distribution of the TA events and determine whether they are statistically
compatible.  In this work we use the flux sampling test proposed by
\citet{Koers:2008ba}.  The starting point is the map of the expected flux
calculated as explained above. One reads off the flux values at positions of
the data events. This gives a set of numbers which we refer to as the ``data
set''. One may say that the cosmic ray events sample the flux map in a
particular way that depends on their space distribution. One then generates a
large number of Monte-Carlo events which are distributed according to the
expected flux and reads off the flux values at their positions.  This gives
the set of flux values which we refer to as the ``MC set''. If the {\em
  angular} distribution of the data and MC events is the same, so must be the
distributions of the flux values in the data and MC sets. These two
distributions may be compared by the parameter-free Kolmogorov-Smirnov (KS)
test.
   
The result of the KS test is the $p$-value which shows whether the data and MC
flux sets are drawn from the same parent distribution. If this $p$-value is
low, the two distributions of flux values are different and, therefore, the
angular distributions of data and MC sets are different.

\subsection{Estimate of Statistical Power of the Flux Sampling Test}
\label{sec:estim-stat-power}

An important characteristic of a statistical test is its ability to
discriminate between two hypotheses, or the statistical power. For the
case at hand, the statistical power is the probability to rule out the
LSS hypothesis at 95\% CL if the cosmic ray distribution is
isotropic. The closer the statistical power to one, the more sensitive
is the test. Knowing the statistical power provides an {\em a priori}
idea of what kind of sensitivity can be reached with the given number
of events.

In general, the statistical power increases with smaller smearing
angles since this improves the contrast in the flux map.  For the same
reason, the statistical power increases with energy (the UHECR
propagation length becomes shorter and the relative contribution of the
local structures is therefore enhanced). Also, the statistical power
increases with the number of events.

We have calculated the statistical power of the flux sampling test in
case of TA for the three energy thresholds of $10$~EeV, $40$~EeV, and
$57$~EeV, and smearing angles varying from $2^\circ$ to $14^\circ$. We
have found that for the actual number of events in the TA data set, the
statistical power is below 50\% for smearing angles $\theta>9^\circ$,
$\theta>3^\circ$ and $\theta>4^\circ$ for the above three energy
thresholds, respectively.

\begin{figure}
  \includegraphics[width=.9\columnwidth]{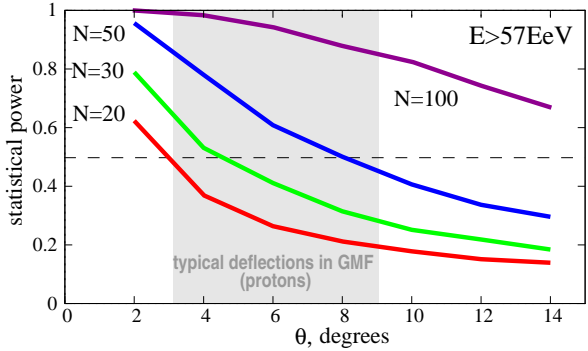}
  \caption{\label{fig:powers} The statistical power of the flux
    sampling test at $E>57$~EeV as a function of the smearing angle.
    Different curves correspond to different number of events, as
    indicated on the plot. The actual number of events in the
    TA data set with $E>57$~EeV is 25. The gray region shows the expected
    range of deflections in the Galactic magnetic field in the case of
    protons. }
\end{figure}

The case $E>57$~EeV is shown in Figure~\ref{fig:powers}.
The various curves in the plot correspond to different number of events (note that
the actual number of events in the TA data set with $E>57$~EeV is 25).
The gray region represents the expected range of deflections in the
Galactic magnetic field in the case of protons.

\subsection{Results}
\label{sec:results}

First, we check the compatibility of the TA event sets with the isotropic
distribution. To this end we generate an isotropic flux map modulated with
the TA exposure. This map is independent of energy and smearing angle. We
then test the compatibility of the TA event set for $E>10$~EeV, $E>40$~EeV, and
$E>57$~EeV with this map. The flux sampling test gives the p-values 0.5, 0.9
and 0.6 for the three data sets, respectively. Thus, at all three energy
thresholds the data appear to be compatible with an isotropic distribution.

\begin{figure}
\begin{picture}(200,369)(0,0)
\put(5,246){\includegraphics[width=.95\columnwidth]{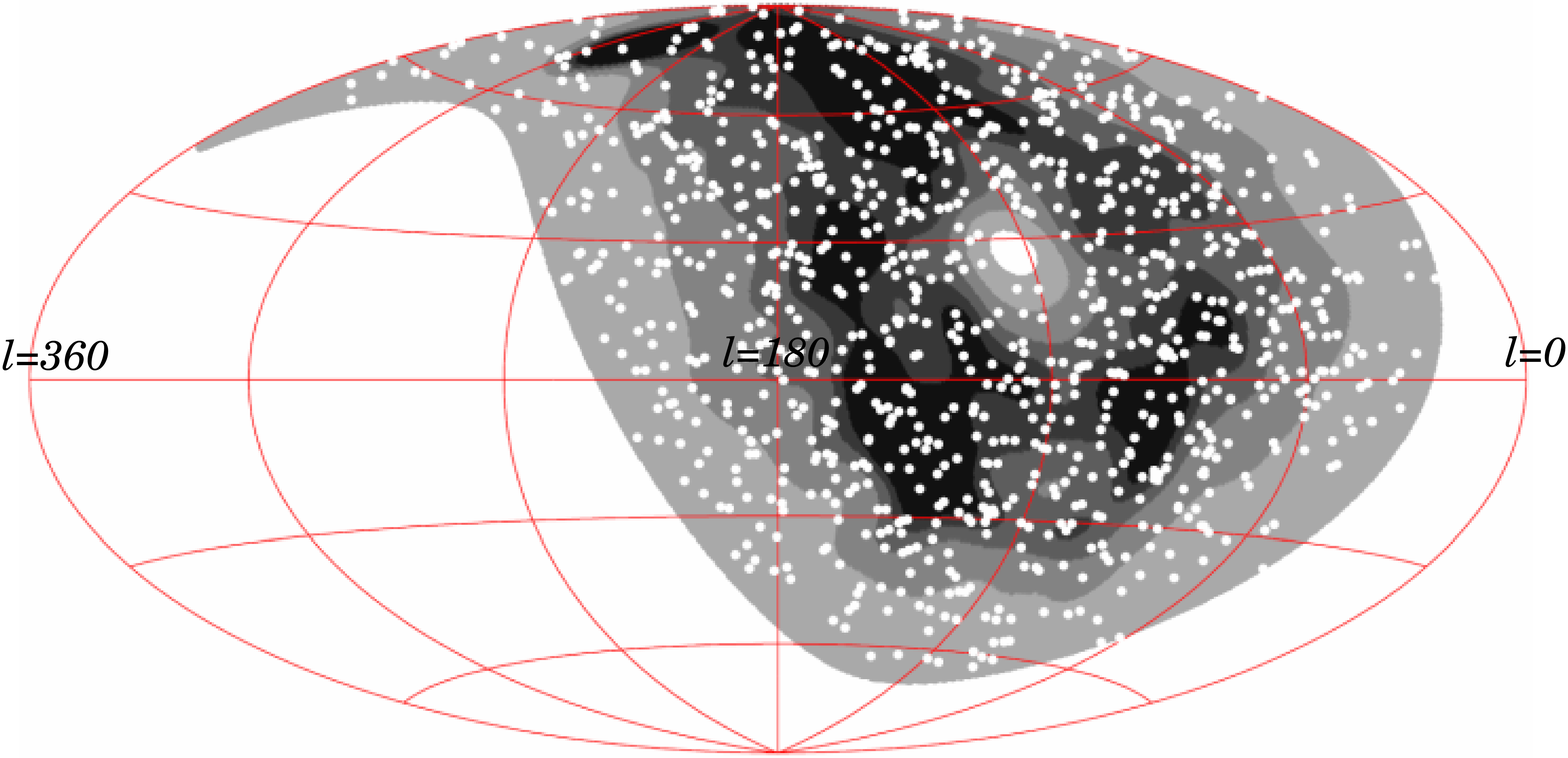}}
\put(5,123){\includegraphics[width=.95\columnwidth]{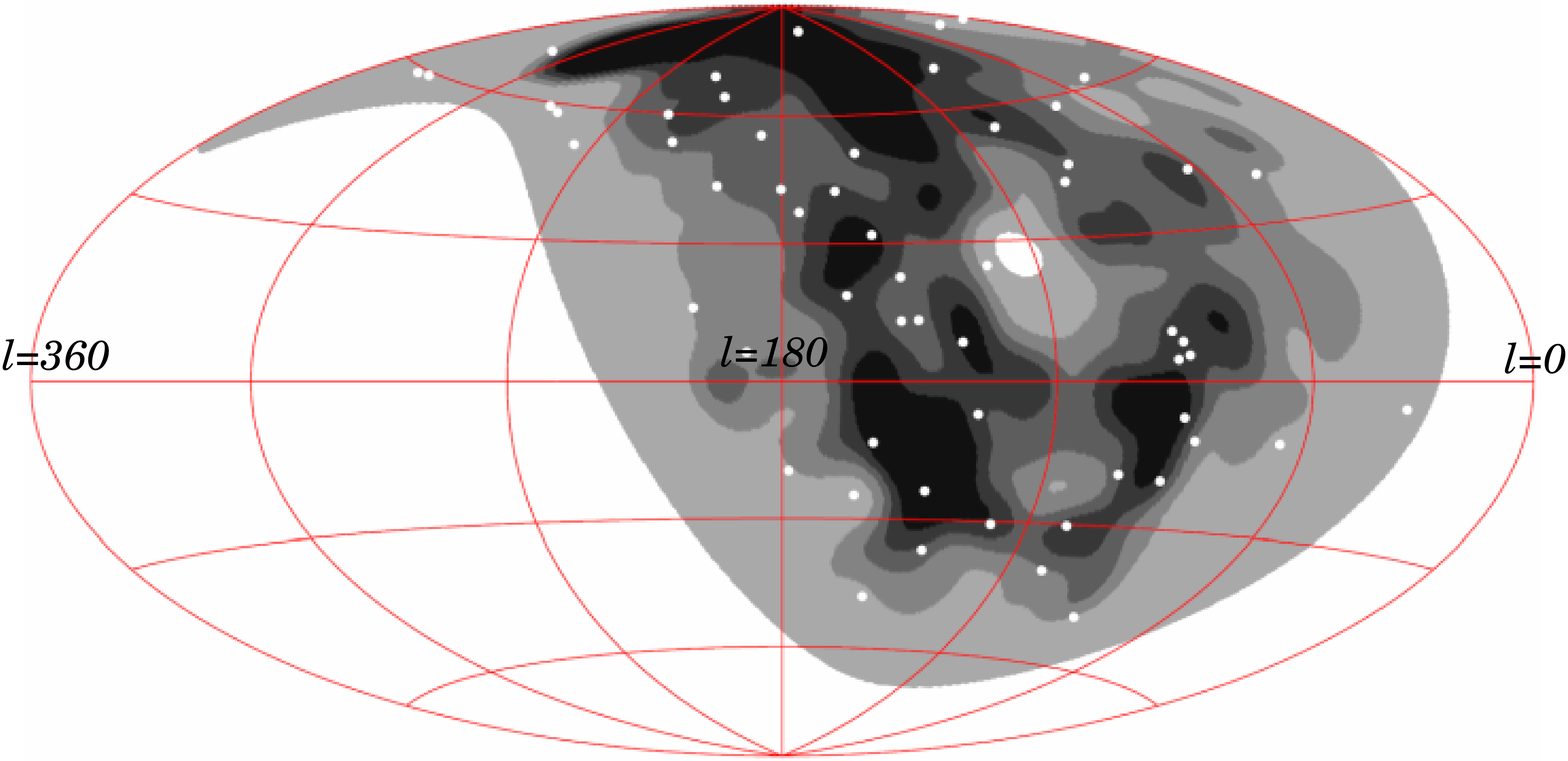}}
\put(5,0){\includegraphics[width=.95\columnwidth]{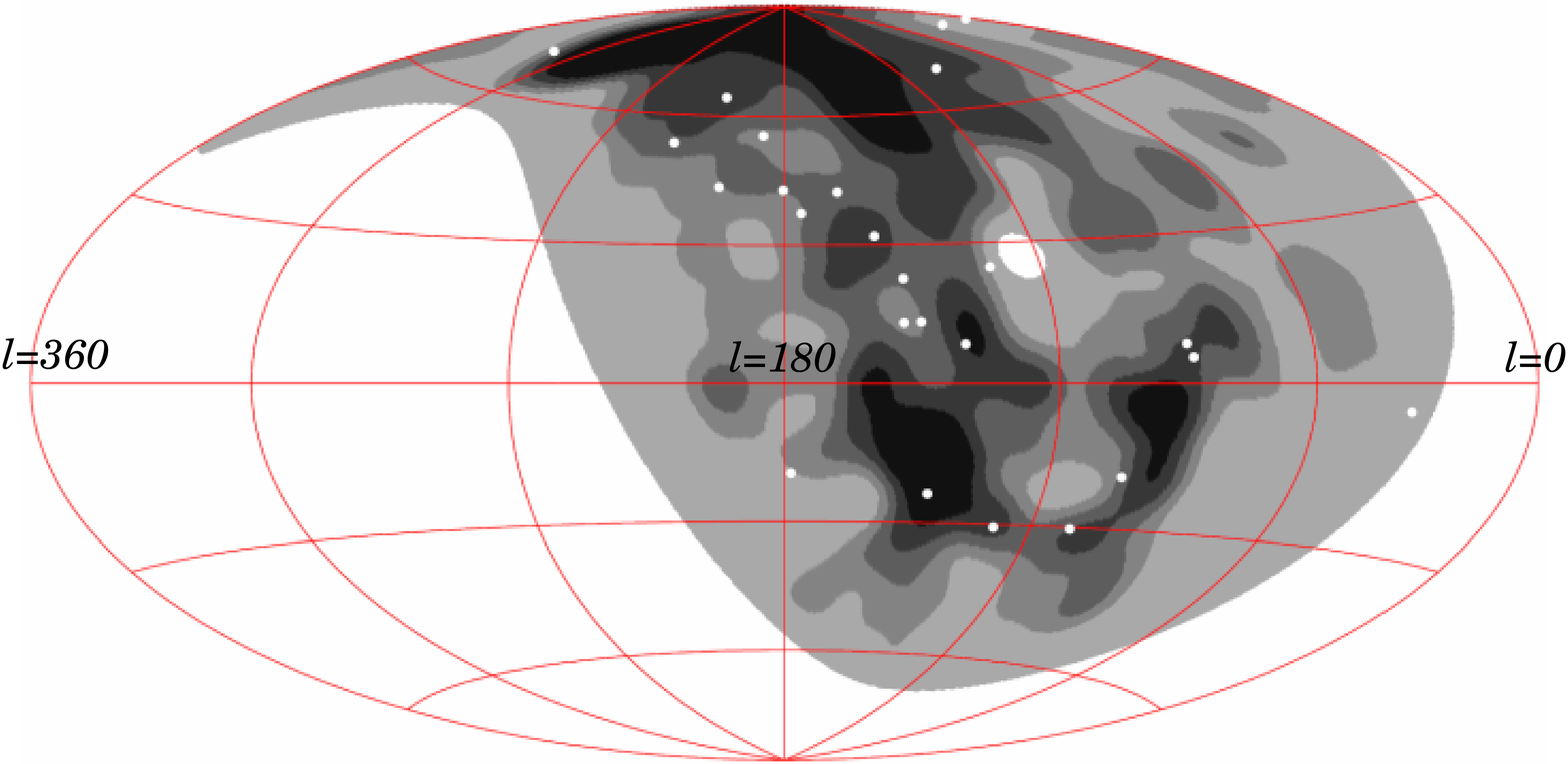}}
\end{picture}
  \caption{ \label{fig:p-skymaps} The skymaps of the expected flux at energy
    thresholds of $10$~EeV, $40$~EeV, and $57$~EeV (from top to bottom) in
    Galactic coordinates with the TA events superimposed (white dots). The
    smearing angle is $6^\circ$. }
\end{figure}

Next, we examine the compatibility of the TA event sets with the LSS
hypothesis. Figure~\ref{fig:p-skymaps} shows the skymaps of the expected flux at
energy thresholds of $10$~EeV, $40$~EeV, and $57$~EeV (top to bottom) and the
smearing angle of $6^\circ$.  The white dots represent the arrival directions
of the TA events. The bands are drawn in the same way as in
Figure~\ref{fig:E57skymap}, i.e., each band integrates to 1/5 of the total
flux. This means, in particular, that if the LSS model were true each band
would contain 1/5 of the total number of events in average. Note that the
configuration of the bands changes with energy because of the energy
dependence of the propagation length.

\begin{figure}
  \includegraphics[width=.9\columnwidth]{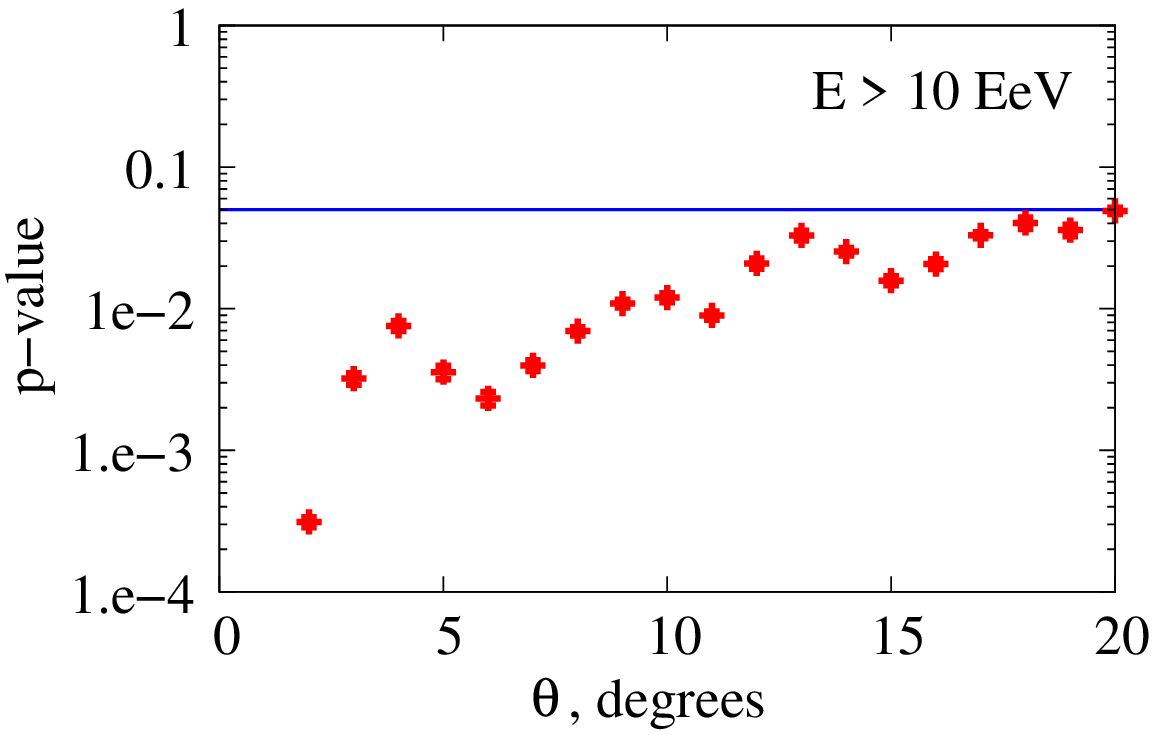}
  \includegraphics[width=.9\columnwidth]{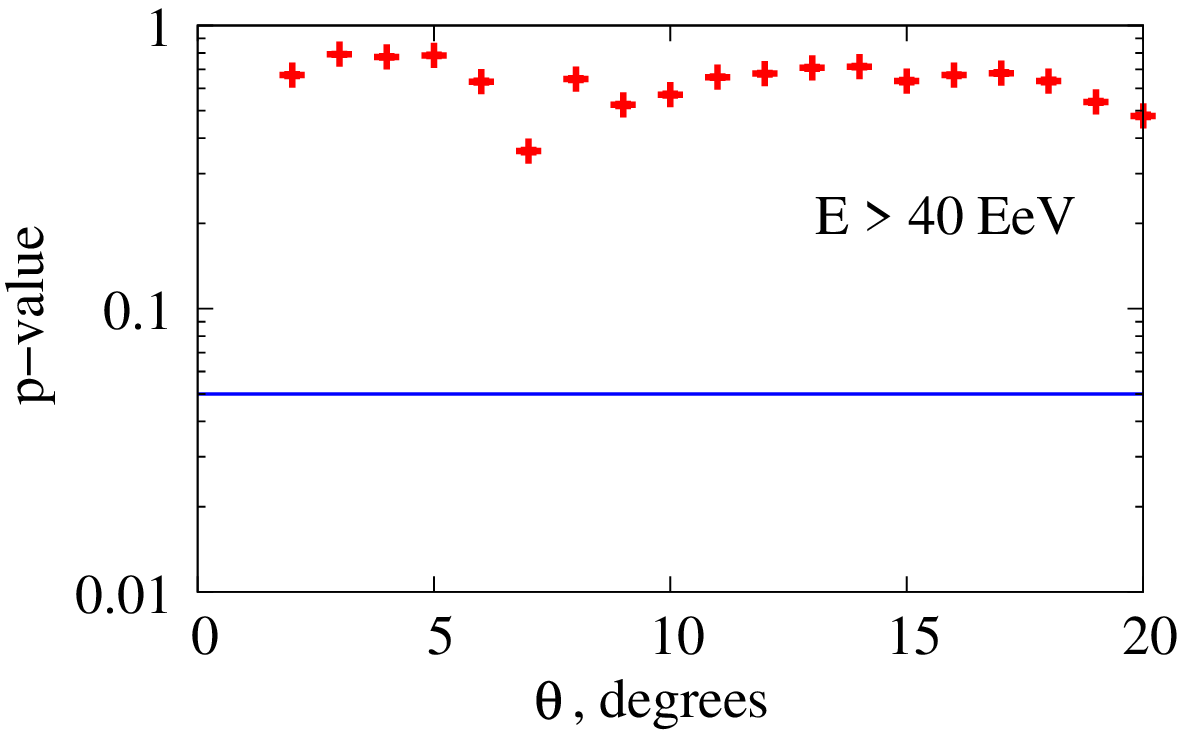}
  \includegraphics[width=.9\columnwidth]{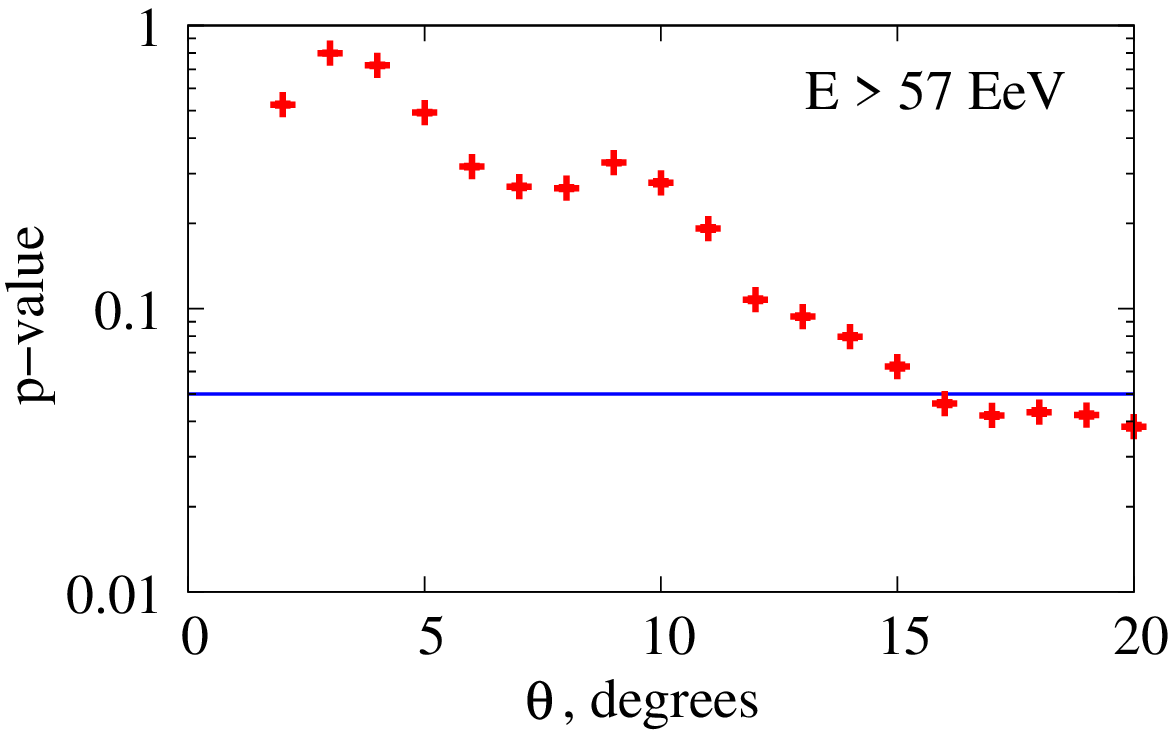}
  \caption{ \label{fig:p-values} The results of the statistical test
    for the compatibility between the data and the LSS hypothesis. The
    p-values (red points) are shown as a function of the smearing
    angle $\theta$. Low p-values indicate incompatibility with the LSS
    model. The horizontal line shows a confidence level of 95\%. The
    three panels correspond to energy thresholds of $10$~EeV,
    $40$~EeV, and $57$~EeV from top to bottom, as indicated on the
    plots.}
\end{figure}

The results of the flux sampling tests are presented in
Figure~\ref{fig:p-values}. The $p$-values are shown as a function of the
smearing angle at energy thresholds of $10$~EeV, $40$~EeV, and $57$~EeV. 
Each point represents the p-value obtained by the flux sampling test at the
corresponding energy threshold and smearing angle.

As one can see from the plots, for $E>40$~EeV and $E>57$~EeV the data
are compatible with the structure hypothesis at the 95\% C.L. The
decrease of the p-values slightly below the 95\% C.L.  in the case
$E>57$~EeV cannot be assigned a real significance in view of the
penalty factors for trials (e.g., three energy thresholds). 

Although large smearing angles do not have a straightforward physical
interpretation in view of the Gaussian approximation used, we have
investigated the behavior of the p-values corresponding to the case $E>57$~EeV
for larger smearing angles and found that it fluctuates around $p\simeq 0.05$
for angles as large as $\theta \sim 50^\circ$ and then goes to 
$p\simeq 1$. Such behavior may arise because the flux map for $E>57$~EeV remains
anisotropic even for very large smearing angles. 

At the energy threshold of $E>10$~EeV the situation is somewhat
different. The data are incompatible with the structure model up to
angles of order $20^\circ$. In view of the large deflections in
magnetic fields at low energies, such behavior is expected. One should
be careful, however, with the interpretation of this result.  First,
Figure~\ref{fig:p-values} does not include the penalty for the number
of trials. Second, at $E>10$~EeV the uncertainties in the flux
calculation due to the choice of the model parameters (in particular, the
injection index and the evolution parameter) are the largest. Finally,
if the smearing angle is attributed to deflections in the magnetic
fields, the dominant contribution is likely to come from the regular
component of the Galactic magnetic field, as discussed in the next
section. Such large and regular deflections require a more accurate
modeling, which we attempt in the next section.

\subsection{Accounting for the Galactic Magnetic Field}
\label{sec:acco-galact-magn}

The deviation from the structure model at $E>10$~EeV and small
smearing angles is an indication that magnetic field deflections play
an important role in the distribution of the UHECR arrival
directions. In general, several contributions to the deflections are
expected. First, there are deflections produced by intergalactic
magnetic fields. These fields are known quite poorly. They are usually
thought to obey the upper bound of $B\lsim 10^{-9}$~G with a
correlation length, $l\lsim 1$~Mpc \citep{Kronberg:1993vk}.  With these
parameters, a proton of energy $10$~EeV coming from $50$~Mpc would
be deflected by $\sim 20^\circ$. However, there are indications that
the extragalactic magnetic fields may be several orders of magnitude
smaller \citep{Dolag:2004kp} than the upper bound.

Second, UHECRs are deflected in the regular component of the Galactic Magnetic
Field (GMF). The regular GMF is known much better than extragalactic
fields. It can be inferred, e.g., from the Faraday rotation measures of
Galactic and extragalactic radio sources.  According to recent studies, a
typical deflection of a $10$~EeV proton would be $20-40^\circ$
\citep{Pshirkov:2011um}. This is comparable or larger than the deflection in
the extragalactic field.

Finally, the Galactic field has a random component. Although the
amplitude of this component is a few times larger than the regular
one, its contribution into the deflections is subdominant (or at most
comparable) to that of the regular component \citep{Tinyakov:2004pw}
due to its random character.

From this discussion it is clear that, most likely, the regular part
of the magnetic field provides the dominant contribution into the UHECR
deflections. At low energies when the magnitude of the deflections
becomes large, Gaussian smearing is not a good approximation for such
deflections. They have
to be taken into account explicitly.

In order to see whether the deflections in the regular GMF can be a reason for
the discrepancy between the data and the LSS model we have repeated the
analysis of Section~\ref{sec:results} including the regular GMF. The presence of
the regular magnetic field is taken into account by modifying the expected
flux distribution. The smearing angle remains a free parameter; it accounts
for random deflections in the extragalactic fields and in the random component
of the GMF. The statistical test itself remains unchanged.

We adopt the recent GMF model by \citet{Pshirkov:2011um}. This model has been
obtained by fitting the GMF model parameters to the latest catalog of the
Faraday rotation measures of extragalactic sources. In addition to the 
disk field, this model also contains a toroidal halo field.

Although the fits to the Faraday rotation measures constrain the parameters of
the GMF, some combinations of these parameters are constrained rather
poorly. In particular, the magnitude of the halo field is degenerate with the
halo height above the Galactic disk: making the halo field stronger and
simultaneously higher above the disk does not strongly affect the rotation
measures. Thus, there remains some freedom in the choice of the GMF
parameters. The question is whether this freedom can be used to bring the
arrival directions of UHECR into accord with the LSS hypothesis without
contradicting the Faraday rotation data.

\begin{figure}
\begin{center}
\begin{picture}(210,280)(0,0)
\put(0,150){\includegraphics[width=.95\columnwidth]{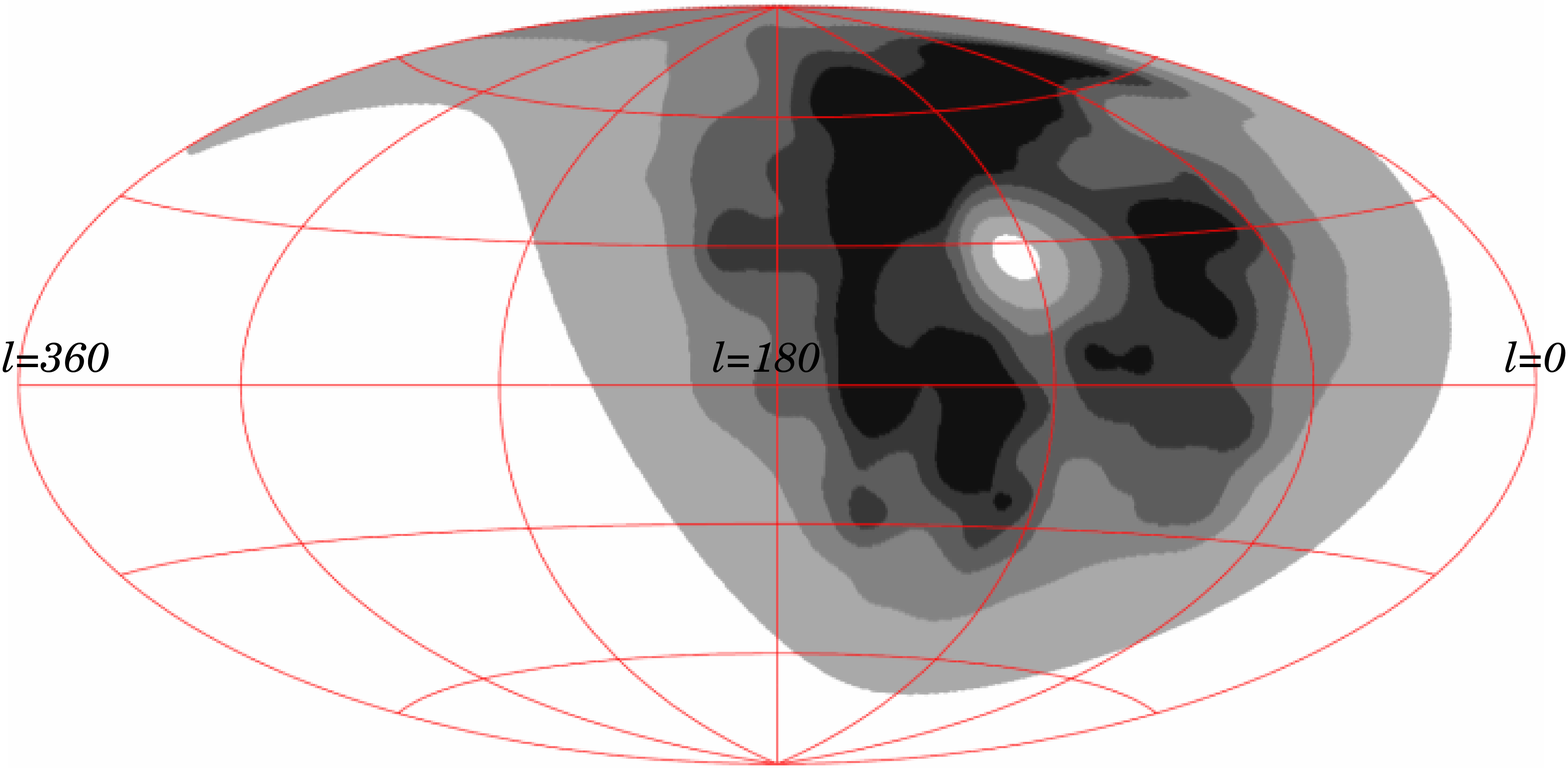}}
\put(0,0){\includegraphics[width=.9\columnwidth]{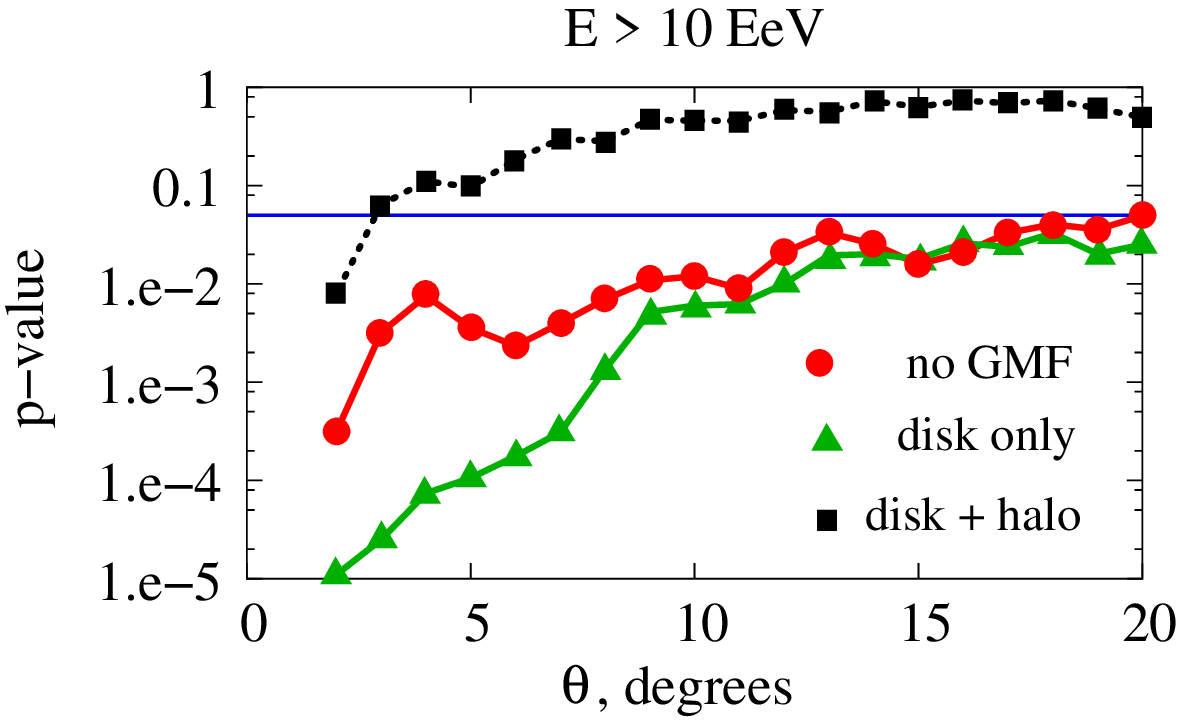}}
\end{picture}
\end{center}
  \caption{ \label{fig:GMFincluded} {\em Upper panel:} The sky map of the
    expected flux for $E>10$~EeV and smearing angle $6^\circ$ taking into account
    the GMF (Galactic coordinates). The parameters of GMF are as follows: 
    the magnitude of the halo is 4~$\mu$G and the thickness of the 
    halo is 1.5~kpc. Note the absence of
    overdensity in the direction of the Virgo cluster.  {\em Lower panel:} The
    result of the statistical test of the compatibility between the TA event
    set with $E>10$~EeV and the LSS hypothesis for different models of GMF: no
    magnetic field (circles), disk component only (triangles), both disk and
    halo components (squares). The horizontal line shows the confidence level
    of 95\%. Low p-values indicate incompatibility. }
\end{figure}
We have found that the compatibility with the LSS model cannot be reached
without the halo field. When the halo is included, the compatibility with the
LSS model is possible, although the required halo field is rather strong (but
still compatible with the data on the Faraday rotation measures). 

An example of the flux map with the GMF included is shown in the upper panel
of Figure~\ref{fig:GMFincluded}. The flux distribution is calculated for the
case $E>10$~EeV and smearing angle of $6^\circ$. The magnetic field
parameters are as follows: the magnitude of the halo 4~$\mu$G and the
thickness of the halo is 1.5~kpc. Note that after the inclusion of GMF the
Virgo region has moved away from the TA field of view, and the expected flux
distribution has become closer to the uniform one.

The results of the flux sampling test of the LSS model with the regular GMF
included are shown in the lower panel of Figure~\ref{fig:GMFincluded}. Black
squares represent the p-values in the case of the GMF with the parameters
described above. For comparison, red circles show the p-values in the absence
of GMF (the same as the upper panel of Figure~\ref{fig:p-values}), while green
triangles represent the case of GMF with the disk component only. One can see
that the regular GMF can produce deflections that make the data for $E>10$~EeV
compatible with the LSS model for all but the smallest smearing angles.  Thus,
the discrepancy between the LSS hypothesis and the TA data with $E>10$~EeV can,
in principle, be explained by the deflections in the regular GMF.

\section{Summary and conclusions}
\label{sec:conclusions}

In this paper we present a search for anisotropy in the TA data
collected over the period of about 40 months, which is the largest
UHECR data set to date in the Northern hemisphere. The main focus of
this paper is on checking the existing claims: small-scale clustering,
correlation with nearby AGNs, and correlation with the LSS.

The results are summarized as follows. 
\begin{itemize}

\item The TA data show no clustering of the UHECR events at small scales,
  neither at the angular scale of $2.5^\circ$ in the set with $E>40$~EeV as
  reported by the AGASA experiment, nor at any angular scale from 0 to 40
  degrees in the data sets with $E>10$~EeV, $E>40$~EeV, and $E>57$~EeV. There
  is a hint of grouping of events at angular scales of $20-30^\circ$ at
  the highest energies, however the statistical significance of this feature is
  insufficient for a definite conclusion.

\item There is no statistically significant correlation of the TA data with
  $E>57$~EeV with the positions of nearby AGNs from the VCV catalog using the
  parameters reported by the PAO (angular scale of $3.1^\circ$ and redshift cut in
  the VCV catalog $z\leq 0.018$). Out of 25 observed events with $E>57$~EeV, 11
  have been found to correlate with positions of nearby AGNs, while 5.9 are
  expected in average from random coincidences (chance probability of 2\%).

\item The TA event sets with $E>10$~EeV, $E>40$~EeV, and $E>57$~EeV appear
  compatible with a uniform distribution according to the flux sampling
  test. The sets with $E>40$~EeV and $E>57$~EeV are also compatible, at 95\%
  CL, with a model which assumes that sources follow the large-scale
  structure of the Universe (LSS model). The set with $E>10$~EeV is not
  compatible, at 95\% CL, with the LSS model unless the deflections of 
  these UHECR exceeds $20^\circ$.

\item The set with $E>10$~EeV can be made compatible with the LSS model, at
  smearing angles larger than $\sim 3^\circ$, by including the effect of the
  regular component of the Galactic magnetic field and assuming a realistic
  model for the latter. The smearing angle in this case represents the
  deflections in the random Galactic and extragalactic fields. 

\end{itemize}

From the analysis presented, one concludes that there is no apparent deviation
from isotropy in the present TA data. At high energies, this may be merely due
to an insufficient number of events. However, if this tendency persists at
several times larger statistics, it will be difficult to reconcile with the
proton composition of UHECR regardless of the source nature: if the sources
within the GZK volume are numerous, they must follow the (inhomogeneous)
matter distribution. If the source density is small so that there are only a
few within the GZK volume, this very fact will produce anisotropy.

At lower energies the deflections are expected to be large even for
protons, which makes the distribution of the events more
isotropic. However, the number of events is much larger as well, so
that even small deviations from isotropy may become detectable as the
statistics increases.  The fact that for $E>10$~EeV the distribution
of the events is not compatible with the LSS model without assuming a
large ($\gsim 20^\circ$) smearing angle may indicate that we are
observing the first manifestation of UHECR deflections in the Galactic
magnetic fields. The possibility to reconcile the observed UHECR
distribution with the LSS model by correcting for the deflections in
the realistic model of GMF is in accord with this interpretation.

\section{Acknowledgements}
The Telescope Array experiment is supported by the Japan Society for
the Promotion of Science through Grants-in-Aid for Scientific Research
on Specially Promoted Research (21000002) ``Extreme Phenomena in the
Universe Explored by Highest Energy Cosmic Rays'' and for Scientific
Research (S) (19104006), and the Inter-University Research Program of
the Institute for Cosmic Ray Research; by the U.S. National Science
Foundation awards PHY-0307098, PHY-0601915, PHY-0703893, PHY-0758342,
PHY-0848320, PHY-1069280, and PHY-1069286 (Utah) and PHY-0649681
(Rutgers); by the National Research Foundation of Korea (2006-0050031,
2007-0056005, 2007-0093860, 2010-0011378, 2010-0028071, 2011-0002617,
R32-10130); by the Russian Academy of Sciences, RFBR grants
10-02-01406a and 11-02-01528a (INR), IISN project No. 4.4509.10 and
Belgian Science Policy under IUAP VI/11 (ULB); by the Grant-in-Aid for
the Scientific Research (S) No. 19104006 by the Japan Society for the
Promotion of Science. The foundations of Dr. Ezekiel R. and Edna
Wattis Dumke, Willard L. Eccles and the George S. and Dolores Dore
Eccles all helped with generous donations.  The State of Utah
supported the project through its Economic Development Board, and the
University of Utah through the Office of the Vice President for
Research.  The experimental site became available through the
cooperation of the Utah School and Institutional Trust Lands
Administration (SITLA), U.S.~Bureau of Land Management and the
U.S.~Air Force.  We also wish to thank the people and the officials of
Millard County, Utah, for their steadfast and warm support.  We
gratefully acknowledge the contributions from the technical staffs of
our home institutions and the University of Utah Center for High
Performance Computing (CHPC).

\section{Appendix: List of events with $E>57$~EeV}
\label{sec:evlist}

In this Appendix we present the list of events with energy $E>57$~EeV and
zenith angle $\theta < 45^\circ$ that have been recorded by the surface
detector of the Telescope Array from May~11, 2008 to September~15,
2011. During this period, 25 such events were observed. Table~\ref{tab:events}
shows the arrival date and time of these events, the zenith angle $\theta$,
energy in units of EeV, and Galactic coordinates $l$ and $b$ in degrees. The
angular resolution of these events is  $\sim1.5^\circ$, while the
energy resolution is better than $20\%$.

\begin{table}
\begin{center}
\begin{tabular}{lllll}
\hline
date \& time (UTC)~& $\theta$ (deg)& $E$ (EeV)& $l$ (deg)~& $b$ (deg)\\
\hline
2008-06-25 19:45:52 & 32.8 & 82.6 & 178.6 & -19.4 \\
2008-07-15 05:26:31 & 34.4 & 57.7 & 90.5 & ~8.0 \\
2008-08-10 12:45:04 & 38.0 & 122.6 & 102.7 & -19.2 \\
2008-11-08 14:30:41 & 15.5 & 60.0 & 198.0 & ~43.1 \\
2008-12-30 10:49:32 & 4.5 & 59.7 & 187.0 & ~55.3 \\
2009-01-22 22:54:22 & 31.3 & 58.0 & 89.3 & ~5.2 \\
2009-03-28 04:36:08 & 34.2 & 81.2 & 152.8 & ~22.5 \\
2009-03-29 03:43:34 & 20.7 & 75.0 & 158.1 & ~31.9 \\
2009-05-19 02:19:52 & 42.5 & 64.6 & 25.8 & ~77.3 \\
2009-09-19 08:45:52 & 34.7 & 62.0 & 140.5 & ~8.4 \\
2010-01-08 07:17:31 & 19.5 & 57.5 & 175.6 & ~37.2 \\
2010-01-21 03:53:51 & 23.4 & 61.2 & 149.8 & ~13.1 \\
2010-02-22 07:10:34 & 14.5 & 63.5 & 165.7 & ~42.0 \\
2010-08-29 21:20:45 & 36.5 & 69.9 & 180.3 & ~42.4 \\
2010-08-30 20:50:45 & 20.0 & 93.3 & 98.3 & ~69.7 \\
2010-09-19 07:05:00 & 23.6 & 66.8 & 129.1 & -30.6 \\
2010-09-21 20:37:06 & 21.1 & 163.0 & 2.8 & ~76.0 \\
2011-01-05 00:56:23 & 9.3 & 67.4 & 110.0 & -30.4 \\
2011-02-28 16:16:26 & 39.3 & 137.6 & 35.5 & -5.0 \\
2011-04-17 20:20:29 & 34.2 & 74.7 & 153.7 & ~12.9 \\
2011-07-13 19:12:34 & 42.6 & 65.6 & 132.1 & ~24.7 \\
2011-07-22 22:15:41 & 11.6 & 62.2 & 204.5 & ~64.6 \\
2011-07-24 23:17:22 & 36.3 & 61.8 & 316.5 & ~69.5 \\
2011-07-28 15:21:08 & 19.6 & 89.0 & 147.0 & -23.7 \\
2011-08-28 21:14:19 & 31.6 & 63.3 & 215.6 & ~53.2 \\
\hline
\end{tabular}
\end{center}
\caption{ \label{tab:events}
List of Telescope Array events with $E\ge 57$~EeV and zenith angle
$\theta <45^\circ$ recorded from May~11,
2008 to September~15, 2011.
}
\end{table}

\vspace{15cm}



\bibliographystyle{apj}                       
\bibliography{apj-jour,uhe,GMF}

\begin{thebibliography}{46}
\expandafter\ifx\csname natexlab\endcsname\relax\def\natexlab#1{#1}\fi



\bibitem[{Abbasi {et~al.}(2006)}]{Abbasi:2005qy}
Abbasi, R., {et~al.} 2006, Astrophys. J., 636, 680

\bibitem[{Abbasi {et~al.}(2008{\natexlab{a}})}]{Abbasi:2007sv}
---. 2008{\natexlab{a}}, Phys. Rev. Lett., 100, 101101

\bibitem[{Abbasi {et~al.}(2008{\natexlab{b}})}]{Abbasi:2008md}
---. 2008{\natexlab{b}}, Astropart. Phys., 30, 175

\bibitem[{Abbasi {et~al.}(2010{\natexlab{a}})}]{Abbasi:2010xt}
---. 2010{\natexlab{a}}, Astrophys. J. Lett., 713, L64

\bibitem[{Abbasi {et~al.}(2010{\natexlab{b}})}]{Abbasi:2009nf}
---. 2010{\natexlab{b}}, Phys.Rev.Lett., 104, 161101

\bibitem[{Abraham {et~al.}(2007)}]{Cronin:2007zz}
Abraham, J., {et~al.} 2007, Science, 318, 938

\bibitem[{Abraham {et~al.}(2008{\natexlab{a}})}]{Abraham:2007si}
---. 2008{\natexlab{a}}, Astropart. Phys., 29, 188

\bibitem[{Abraham {et~al.}(2008{\natexlab{b}})}]{Abraham:2008ru}
---. 2008{\natexlab{b}}, Phys. Rev. Lett., 101, 061101

\bibitem[{Abraham {et~al.}(2010)}]{Abraham:2010yv}
---. 2010, Phys.Rev.Lett., 104, 091101

\bibitem[{Abreu {et~al.}(2010)}]{Abreu:2010zzj}
Abreu, P., {et~al.} 2010, Astropart. Phys., 34, 314

\bibitem[{Abu-Zayyad {et~al.}(2012{\natexlab{a}})Abu-Zayyad, Aida, Allen,
  Anderson, Azuma, {et~al.}}]{AbuZayyad:2012ru}
Abu-Zayyad, T., Aida, R., Allen, M., {et~al.} 2012{\natexlab{a}},
  arXiv:1205.5067

\bibitem[{Abu-Zayyad {et~al.}(2012{\natexlab{b}})Abu-Zayyad, Aida, Allen,
  Anderson, Azuma, {et~al.}}]{AbuZayyad:2012kk}
---. 2012{\natexlab{b}}, arXiv:1201.4964

\bibitem[{Bird {et~al.}(1999)}]{Bird:1998nu}
Bird, D.~J., {et~al.} 1999, Astrophys. J., 511, 739

\bibitem[{Dolag {et~al.}(2005)Dolag, Grasso, Springel, \&
  Tkachev}]{Dolag:2004kp}
Dolag, K., Grasso, D., Springel, V., \& Tkachev, I. 2005, JCAP, 0501, 009

\bibitem[{Dubovsky {et~al.}(2000)Dubovsky, Tinyakov, \&
  Tkachev}]{Dubovsky:2000gv}
Dubovsky, S.~L., Tinyakov, P.~G., \& Tkachev, I.~I. 2000, Phys. Rev. Lett., 85,
  1154

\bibitem[{Gelmini {et~al.}(2007)Gelmini, Kalashev, \& Semikoz}]{Gelmini:2007jy}
Gelmini, G.~B., Kalashev, O.~E., \& Semikoz, D.~V. 2007, JCAP, 0711, 002

\bibitem[{Giacinti {et~al.}(2010)Giacinti, Kachelriess, Semikoz, \&
  Sigl}]{Giacinti:2010dk}
Giacinti, G., Kachelriess, M., Semikoz, D., \& Sigl, G. 2010, JCAP, 1008, 036

\bibitem[{Glushkov(2001)}]{Glushkov:2001jm}
Glushkov, A. 2001, JETP Lett., 73, 313

\bibitem[{Glushkov \& Pravdin(2001)}]{Glushkov:2001kb}
Glushkov, A., \& Pravdin, M. 2001, J.Exp.Theor.Phys., 92, 887

\bibitem[{Gorbunov {et~al.}(2004)Gorbunov, Tinyakov, Tkachev, \&
  Troitsky}]{Gorbunov:2004bs}
Gorbunov, D.~S., Tinyakov, P.~G., Tkachev, I.~I., \& Troitsky, S.~V. 2004, JETP
  Lett., 80, 145

\bibitem[{Gorbunov {et~al.}(2006)Gorbunov, Tinyakov, Tkachev, \&
  Troitsky}]{Gorbunov:2005fi}
---. 2006, JCAP, 0601, 025

\bibitem[{Greisen(1966)}]{Greisen:1966jv}
Greisen, K. 1966, Phys. Rev. Lett., 16, 748

\bibitem[{{Han} {et~al.}(2006){Han}, {Manchester}, {Lyne}, {Qiao}, \& {van
  Straten}}]{Han2006}
{Han}, J.~L., {Manchester}, R.~N., {Lyne}, A.~G., {Qiao}, G.~J., \& {van
  Straten}, W. 2006, \apj, 642, 868

\bibitem[{Hayashida {et~al.}(1996)}]{Hayashida:1996bc}
Hayashida, N., {et~al.} 1996, Phys. Rev. Lett., 77, 1000

\bibitem[{Kachelriess \& Semikoz(2005)}]{Kachelriess:2004pc}
Kachelriess, M., \& Semikoz, D. 2005, Astropart. Phys., 23, 486

\bibitem[{Kashti \& Waxman(2008)}]{Kashti:2008bw}
Kashti, T., \& Waxman, E. 2008, JCAP, 0805, 006

\bibitem[{Kewley {et~al.}(1996)Kewley, Clay, \& Dawson}]{Kewley:1996zt}
Kewley, L.~J., Clay, R.~W., \& Dawson, B.~R. 1996, Astropart. Phys., 5, 69

\bibitem[{{Koers} \& {Tinyakov}(2009{\natexlab{a}})}]{Koers:2009pd}
{Koers}, H. B.~J., \& {Tinyakov}, P. 2009{\natexlab{a}}, Mon. Not. Roy. Astron. Soc., 399,
  1005

\bibitem[{Koers \& Tinyakov(2009{\natexlab{b}})}]{Koers:2008ba}
---. 2009{\natexlab{b}}, JCAP, 0904, 003

\bibitem[{Kronberg(1994)}]{Kronberg:1993vk}
Kronberg, P.~P. 1994, Rept.Prog.Phys., 57, 325

\bibitem[{{Lynden-Bell}(1971)}]{Lynden-Bell:1971}
{Lynden-Bell}, D. 1971, \mnras, 155, 95

\bibitem[{Pshirkov {et~al.}(2011)Pshirkov, Tinyakov, Kronberg, \&
  Newton-McGee}]{Pshirkov:2011um}
Pshirkov, M., Tinyakov, P., Kronberg, P., \& Newton-McGee, K. 2011,
  Astrophys.J., 738, 192

\bibitem[{Stanev {et~al.}(1995)Stanev, Biermann, Lloyd-Evans, Rachen, \&
  Watson}]{Stanev:1995my}
Stanev, T., Biermann, P.~L., Lloyd-Evans, J., Rachen, J.~P., \& Watson, A.~A.
  1995, Phys. Rev. Lett., 75, 3056

\bibitem[{{Sun} {et~al.}(2008){Sun}, {Reich}, {Waelkens}, \&
  {En{\ss}lin}}]{Sun:2007mx}
{Sun}, X.~H., {Reich}, W., {Waelkens}, A., \& {En{\ss}lin}, T.~A. 2008, \aap,
  477, 573

\bibitem[{Takami {et~al.}(2012)Takami, Inoue, \& Yamamoto}]{Takami:2012uw}
Takami, H., Inoue, S., \& Yamamoto, T. 2012

\bibitem[{Tameda(2010)}]{Tameda:2010-uhecr2010}
Tameda, Y. 2010, prepared for International Symposium on the Recent Progress of
  Ultra-high Energy Cosmic Ray Observation (UHECR2010), Nagoya, JAPAN Dec.10-12

\bibitem[{Tinyakov \& Tkachev(2001)}]{Tinyakov:2001ic}
Tinyakov, P.~G., \& Tkachev, I.~I. 2001, JETP Lett., 74, 1

\bibitem[{Tinyakov \& Tkachev(2004)}]{Tinyakov:2004bb}
---. 2004, Phys. Rev., D69, 128301

\bibitem[{Tinyakov \& Tkachev(2005)}]{Tinyakov:2004pw}
---. 2005, Astropart. Phys., 24, 32

\bibitem[{Tokuno {et~al.}(2012)Tokuno, Tameda, Takeda, Kadota, Ikeda,
  {et~al.}}]{Tokuno:2012mi}
Tokuno, H., Tameda, Y., Takeda, M., {et~al.} 2012, NIM-A, 676, 54

\bibitem[{Veron-Cetty \& Veron(2006)}]{VeronCetty:2006zz}
Veron-Cetty, M.-P., \& Veron, P. 2006, Astron.Astrophys., 455, 773

\bibitem[{Yoshiguchi {et~al.}(2004)Yoshiguchi, Nagataki, \&
  Sato}]{Yoshiguchi:2004np}
Yoshiguchi, H., Nagataki, S., \& Sato, K. 2004, Astrophys.J., 614, 43

\bibitem[{Yoshiguchi {et~al.}(2003)Yoshiguchi, Nagataki, Tsubaki, \&
  Sato}]{Yoshiguchi:2002rb}
Yoshiguchi, H., Nagataki, S., Tsubaki, S., \& Sato, K. 2003, Astrophys.J., 586,
  1211

\bibitem[{Zatsepin \& Kuzmin(1966)}]{Zatsepin:1966jv}
Zatsepin, G.~T., \& Kuzmin, V.~A. 1966, JETP Lett., 4, 78

\end{thebibliography}
\end{document}